%% file: main.tex
\documentclass[twocolumn]{article}
\usepackage[margin=2cm]{geometry} 
\usepackage{authblk} 
\usepackage{graphicx} 
\usepackage{biblatex} 
\usepackage{subcaption}
\usepackage{physics} 
\usepackage{xcolor} 
\usepackage{hyperref}

\title{Magnetically Confined Mountains on Accreting White Dwarfs}
\author[1]{Pedro H. B. Rossetto}
\author[2]{Manoel F. Sousa}
\author[1]{Diego A. Falceta-Gonçalves}
\affil[1]{Escola de Artes, Ciências e Humanidades, Universidade de São Paulo \\ Rua Arlindo Bettio, CEP 03828-000, 1000 São Paulo, SP, Brazil}
\affil[2]{Instituto de Física de São Carlos, Universidade de São Paulo, Av. Trabalhador São-carlense 400, São Carlos, SP, Brazil}
\date{}

\begin{document}
\twocolumn[{
\maketitle
\begin{abstract}
    \input{Sections/0-abstract}
\end{abstract}
\vspace{2em} 
}]

\section{Introduction}
\input{Sections/1-introduction}

\section{Theory of Magnetostatic Equilibrium}
\input{Sections/2-mountains_theory}

\section{Numerical Methods}
\label{sec:numerical_methods}
\input{Sections/3-numerical_methods}

\section{Magnetically Confined Mountains on White Dwarfs}
\label{sec:hydromag_structure}
\input{Sections/4-hydromag_structure}

\section{Magnetic Burial}
\label{sec:mag_burial}
\input{Sections/5-mag_burial}

\section{Summary and Discussion}
\input{Sections/6-conclusion}

\section*{Acknowledgments}
\input{Sections/acknowledgments.tex}

\printbibliography

\appendix
\section{Effect of the Outer Boundary Condition}
\label{sec:app_bc}
\input{Sections/appendix}
\end{document}

%% file: Sections/0-abstract.tex
The hydromagnetic structure of magnetically confined mountains on accreting white dwarfs is computed, alongside the effects of accretion on the reduction of the star's magnetic field. The equilibrium structure of the mountain is obtained by numerically solving the Grad-Shafranov equation with a self-consistent scheme that enforces magnetic flux freezing. For characteristic system parameters, it is shown that magnetic field lines are significantly deformed and dragged equatorwards by mountains with masses $\sim10^{-3}M_\odot$ that have a characteristic height of $\sim10^5\,\mathrm{cm}$. A mathematical relation is obtained for the maximum mass that a white dwarf can confine magnetically given the star's magnetic field and temperature. It is found that the accretion buries the star's magnetic field by reducing it by up to $10\%$ of its pre-accretion value. The star's magnetic field remains dominantly dipolar, but with added multipolar components.

\,

\textbf{Key words:} white dwarf stars, magnetic fields, accretion, stellar accretion, magnetohydrodynamics, plasma astrophysics

%% file: Sections/1-introduction.tex
Magnetized accretion governs the evolution of several astrophysical objects, from young stellar systems to white dwarfs, neutron stars and black holes. For sufficiently strong large-scale magnetic fields, however, the accretion flow is modified. Instead of reaching the stellar surface through an approximately axisymmetric boundary layer, the inflowing plasma becomes magnetically channelled along field lines, producing localized accretion columns or polar caps \cite{Pringle1972,Elsner1977,Ghosh1978,Ghosh1979,Frank2002}. The interaction between accretion flows and stellar magnetic fields therefore plays a central role in determining the structure of the accreted plasma, the evolution of the magnetic field itself, and the observational properties of magnetized compact objects.

Magnetically channelled accretion is intrinsically a multidimensional magnetohydrodynamic problem. Material deposited on the magnetic poles exerts pressure on the underlying magnetic field, causing the field lines to deform while magnetic tension opposes the lateral spreading of the accumulated plasma. The resulting equilibrium reflects a competition between gravity, gas pressure and magnetic stresses. As accretion proceeds, progressively larger amounts of material can become confined near the magnetic poles, distorting the external magnetic field and partially burying the original dipolar configuration. This process, commonly referred to as \emph{magnetic burial}, has important consequences for the long-term magnetic evolution of compact stars, the geometry of the magnetosphere, and the production of non-axisymmetric mass distributions capable of generating continuous gravitational radiation \cite{payne2004burial,melatos2005gravitational, rossetto2023magnetically,rossetto2025quadrupole}.

The formation of magnetically confined mountains has been investigated mainly in the accreting neutron stars scenario. Early theoretical considerations suggested that the intense magnetic fields of neutron stars could efficiently support localized accumulations of accreted matter against gravity while simultaneously modifying the global magnetic topology \cite{melatos2001hydromagnetic}. \citeauthor{payne2004burial}, \citeyear{payne2004burial} -- hereafter PM04 -- solved the Grad--Shafranov equation self-consistently for an axisymmetric dipolar magnetic field, demonstrating that substantial masses may remain magnetically supported near the stellar poles while producing significant magnetic burial \cite{payne2004burial}. Their calculations showed that the external dipole moment decreases as accretion proceeds, whereas the accumulated matter develops a finite mass quadrupole that may become a persistent source of continuous gravitational waves in rapidly rotating systems \cite{melatos2005gravitational,priymak2011quadrupole, rossetto2025quadrupole}. Subsequent work explored the influence of realistic equations of state, resistive relaxation, sinking of accreted material into the stellar crust, and three-dimensional magnetic topology on the equilibrium structure and long-term evolution of the mountain \cite{wette2010sinking,priymak2011quadrupole}. More recently, increasingly sophisticated numerical calculations have incorporated realistic stellar microphysics and generalized magnetic configurations, leading to improved estimates of magnetic burial, stellar deformation and gravitational-wave emission \cite{fujisawa2022magnetically,rossetto2023magnetically,rossetto2025quadrupole,yeole2025investigating}. 

While the theoretical development of magnetic mountains has focused almost exclusively on neutron stars, the basic physical ingredients are also present in other classes of objects. White dwarfs, for instance, also undergo magnetically channelled accretion over a broad range of astrophysical environments. In interacting binaries, such as magnetic cataclysmic variables, polars (AM Her systems) and intermediate polars (DQ Her systems), matter transferred from a Roche-lobe-filling companion is funnelled by the stellar magnetic field toward one or both magnetic poles, producing localized accretion columns and strong X-ray emission \cite{Cropper1990,Warner1995,Ferrario2015,Mukai2017}. The interaction between the accretion flow and the magnetosphere in these systems has been extensively investigated both observationally and theoretically, establishing magnetic confinement as a fundamental ingredient of accretion onto highly magnetized white dwarfs. Also, strong large-scale magnetic fields are relatively common among white dwarfs. Magnetic field strengths inferred from Zeeman splitting and spectropolarimetric observations span nearly six orders of magnitude, from approximately $10^3$ to $10^9$~G, with the strongest fields preferentially found among magnetic cataclysmic variables \cite{Wickramasinghe2000,Ferrario2015,Ferrario2020}. In polars, where surface magnetic fields frequently exceed several tens of megagauss, the magnetic pressure dominates the dynamics of the inflowing plasma over a substantial fraction of the magnetosphere, forcing the accretion flow to remain tightly confined until it reaches the stellar surface. 

Recent observations further reinforce this picture. The discovery of micronovae on strongly magnetized white dwarfs has provided compelling evidence that thermonuclear burning may occur within localized regions produced by magnetically channelled accretion rather than over the entire stellar surface \cite{Scaringi2022}. These observations indicate that magnetic confinement can remain effective over sufficiently long timescales to accumulate appreciable amounts of material near the magnetic poles before ignition occurs. Although micronovae probe relatively shallow layers of freshly accreted matter, they provide direct observational support for the existence of localized magnetically confined structures on white dwarf surfaces and therefore strengthen the physical motivation for investigating whether larger-scale magnetically supported equilibria may also develop.

Taken together, these observational and theoretical results suggest that the essential ingredients responsible for the formation of magnetic mountains in neutron stars are also present in white dwarfs. Nevertheless, despite the extensive literature devoted to magnetic burial and magnetically confined mountains in neutron stars, the corresponding problem has received remarkably little attention in the context of white dwarfs. Whether white dwarf magnetic fields are capable of sustaining analogous equilibrium structures, the total accreted mass that can be magnetically supported, and how the magnetic topology evolves during the accretion process remain as open questions.

In this paper we investigate the equilibrium structure of magnetically confined accretion mountains on white dwarfs by extending the Grad--Shafranov formalism previously developed for neutron stars to the physical conditions appropriate for white dwarf envelopes. We compute self-consistent magnetohydrostatic equilibria for a range of magnetic field strengths and accreted masses, quantify the resulting magnetic burial and stellar deformation, and examine the dependence of the equilibrium configuration on the adopted physical parameters. By establishing the conditions under which magnetic mountains may form on white dwarfs, this work aims to provide a theoretical framework connecting the well-developed theory of neutron-star magnetic mountains with the increasingly rich phenomenology of magnetized accreting white dwarfs.

%% file: Sections/2-mountains_theory.tex
In this paper, we investigate the equilibrium of accreted matter on a white dwarf's (WD's) magnetic pole. The theory used here has its origins in neutron star physics \cite{payne2004burial,priymak2011quadrupole,rossetto2023magnetically}, but it can be successfully applied to white dwarfs. In this section, we present the main equations that govern the system. In Section \ref{sub-sec:grad-shafranov}, we discuss the equilibrium equation; in Section \ref{sub-sec:flux-freezing}, we explain an integral condition that needs to be enforced for a self-consistent set of equations; and, in Section \ref{sub-sec:boundary-conditions}, we discuss the boundary conditions used.

\subsection{The Grad-Shafranov Equation}
\label{sub-sec:grad-shafranov}

The equations of magnetohydrodynamics (MHD) govern the dynamics of accreted plasma on white dwarfs,
\begin{align}
    \label{mass_cons}
    &\pdv{\rho}{t} + \div{(\rho \mathbf{u})} = 0,
    \\
    \label{mom_eq}
    &\rho\pdv{\mathbf{u}}{t} + \rho (\mathbf{u}\cdot\grad)\mathbf{u}= - \rho\grad{\Phi} - \grad{p} + \frac{(\curl{\mathbf{B}})}{4\pi} \cross \mathbf{B},
    \\
    \label{ind_eq}
    &\pdv{\mathbf{B}}{t} = \curl{(\mathbf{u}\cross\mathbf{B})} + \frac{1}{\sigma}\nabla^2\mathbf{B},
\end{align}
where $\rho$ is the plasmas density, $t$ is time, $\mathbf{u}$ is the plasma velocity, $\Phi$ is the gravitational potential, $p$ is the plasma pressure, $\mathbf{B}$ is the magnetic field, and $\sigma$ is the electrical conductivity. For static equilibrium configurations, $\mathbf{u} \equiv 0$ and $\partial/\partial t \equiv 0$, this makes equation \eqref{mass_cons} identically satisfied and equation \eqref{mom_eq} becomes
\begin{equation}
    \label{equil_mom_eq}
    \rho\grad{\Phi} + \grad{p} - \frac{(\curl{\mathbf{B}})}{4\pi} \cross \mathbf{B} = 0.
\end{equation}
Assuming ideal MHD, where $\sigma\to \infty$, equation \eqref{ind_eq} is also identically satisfied. 

We consider the case of purely poloidal magnetic fields and, hence, we write
\begin{equation}
    \label{mag_field}
    \mathbf{B} = \frac{\grad{\psi}}{r\sin\theta}\cross \hat{\phi}
\end{equation}
where $\psi$ is the magnetic field flux function, and $(r,\theta,\phi)$ are the standard spherical coordinates with $\hat{\phi}$ being the unit vector in the $\phi$ direction. Substituting this form of $\mathbf{B}$ into equation \eqref{equil_mom_eq}, we obtain
\begin{equation}
    \label{phi_p_psi}
    \rho\grad{\Phi} + \grad{p} + \Delta^*\psi\grad{\psi} = 0,
\end{equation}
where $\Delta^*$ is the Grad-Shafranov operator defined by
\begin{equation}
    \label{grad_shaf_op}
    \Delta^* = \frac{1}{4\pi r^2\sin^2\theta}\left[\pdv[2]{r}+\frac{\sin\theta}{r^2}\pdv{\theta}\left(\frac{1}{\sin\theta}\pdv{\theta}\right)\right].
\end{equation}
Considering a barotropic equation of state, $p = p(\rho)$, we can find a function $U$ such that $\grad{U}=\rho^{-1}\grad{p}$; then, equation \eqref{phi_p_psi} implies that $\Phi+U = f(\psi)$, for a function $f(\psi)$.  Substituting these results into equation \eqref{phi_p_psi}, we obtain
\begin{equation}
    \label{grad_shafranov}
    \Delta^*\psi = -\rho f'(\psi),
\end{equation}
which is the Grad-Shafranov equation for plasma equilibrium. As a first approach to the problem, we use an isothermal equation of state, $p = {c_s}^2\rho$ with $c_{\rm s}$ being the sound speed, and we obtain the isothermal Grad-Shafranov equation,
\begin{equation}
    \label{iso_grad_shafranov}
    \Delta^*\psi=-F'(\psi)e^{-(\Phi-\Phi_0)/c_s^2},
\end{equation}
where $F(\psi)=e^{f(\psi)}$ and $\Phi_0$ is a reference potential. The function $F(\psi)$ is determined by the flux-freezing condition of ideal MHD, and an equation for it will be presented in Section \ref{sub-sec:flux-freezing}. For the gravitational potential $\Phi$, we use the surface approximation,
\begin{equation}
    \label{Phi}
    \Phi(r) = \frac{GM_*}{{R_*}^2}r,
\end{equation}
where $G$ is the gravitational constant, $M_*$ is the mass of the star and $R_*$ is the star's radius. This approximation is justified because the scale heights of the system are much smaller than the star's radius (see Section \ref{sec:hydromag_structure}).

\citeauthor{rossetto2023magnetically} \citeyear{rossetto2023magnetically} and \citeyear{rossetto2025quadrupole} showed that the general relativistic corrections in the modelling of magnetically confined mountains on neutron stars yield sizeable corrections, modifying the scaling relations of the magnetic dipole moment by three times and the ellipticity by up to $12\%$  \cite{rossetto2023magnetically, rossetto2025quadrupole}. Therefore, one might ask if these corrections are also important for white dwarfs. A brief quantitative assessment of the importance of the general relativistic corrections can be done by using the compactness factor $\mathcal{C}$, which is given by
\begin{equation}
    \label{compactness}
    \mathcal{C} = \frac{GM_*}{c^2R_*},
\end{equation}
where $c$ is the speed of light in vacuum. A typical value for the compactness of a neutron star is $2\times10^{-1}$, while for a white dwarf is $2\times10^{-4}$. This suggests that relativistic corrections should be substantially smaller for white dwarfs. Furthermore, we have tested the extent of these corrections by running general relativistic simulations of the problem -- using the methods outlined in \cite{rossetto2023magnetically} -- and we have verified that the corrections are indeed small. Using the simulation shown in Figure \ref{fig:psi_contours} as a comparison, the percent error in the normalised magnetic dipole moment was $0.2\%$. Therefore, we choose to use the classical theory in the present paper for ease of presentation and interpretation.

\subsection{Flux-Freezing Constraint}
\label{sub-sec:flux-freezing}

To solve equation \eqref{iso_grad_shafranov}, one needs to provide the function $F(\psi)$. This can be done using a few different approaches. The simplest method is to give an analytic form of $F(\psi)$ \cite{melatos2001hydromagnetic, fujisawa2022magnetically}, however, it is not clear how to impose a physically motivated function in this way. Another approach is to relate the $F(\psi)$ with the mountain's density and then impose a height and profile to the mountain \cite{yeole2025investigating, mukherjee2017revisiting}. Albeit physically motivated, this method might introduce violations to the flux-freezing condition and, therefore, make the equations \eqref{iso_grad_shafranov} and \eqref{Fpsi} not self-consistent.

A physically-motivated and self-consistent approach was introduced by \citeauthor{payne2004burial}, \citeyear{payne2004burial} \cite{payne2004burial}, where $F(\psi)$ is related to the density and the mass-flux ratio $\dd{M}/\dd{\psi}$ is set in a way that mass does not cross magnetic flux surfaces and, therefore, satisfies the flux-freezing condition. This is the approach we adopt in this paper. Using the isothermal equation of state and the definitions presented in Section \ref{sub-sec:grad-shafranov}, the density can be written as 
\begin{equation}
    \label{rho}
    \rho = \frac{F(\psi)}{{c_s}^2}e^{-(\Phi-\Phi_0)/c_s^2}.
\end{equation}
Then, the mass per flux is
\begin{equation}
    \label{dmdpsi}
    \frac{dM}{d\psi} = 2\pi\int_{\mathcal{C}} r\sin\theta|\grad\psi|^{-1}\rho\left[r(s),\theta(s)\right] \dd{s},
\end{equation}
where $\mathcal{C}$ is the line of constant $\psi$ -- i.e., the magnetic field line (see equation \eqref{mag_field}) -- and $s$ is the parameter along the line. Using equation \eqref{rho} in equation \eqref{dmdpsi}, we can find $F(\psi)$ to be
\begin{equation}
    \label{Fpsi}
    F(\psi) = \frac{{c_s}^2}{2\pi}\frac{dM}{d\psi}\left\{\int_\mathcal{C}ds\, r\sin\theta|\grad{\psi}|^{-1}e^{-(\Phi - \Phi_0)/{c_s}^2}\right\}^{-1}.
\end{equation}
Equations \eqref{iso_grad_shafranov} and \eqref{Fpsi} form the self-consistent set of equations that we need to solve for the static equilibrium of the accreted matter on the pole of white dwarfs.

For the right-hand side of equation \eqref{Fpsi} to be completely determined, we need to know the mass-flux distribution $\dd{M}/\dd{\psi}$. This function depends on the complex accretion history of the system, the geometry of the accretion disk, and the magnetosphere-disk interface; all of which lie beyond the scope of this paper. We choose a functional dependency for the mass-flux distribution that retains the qualitative characteristics of the accretion dynamics; i.e., we concentrate the mass on the polar cap -- within a flux surface $\psi_{\rm a}$ -- and let the distribution have a smooth tail. The flux line $\psi_{\rm a}$ is to be thought of as the magnetic field line that touches the inner edge of the accretion disk, $\psi_{\rm a} = \psi(R_{\rm m},\pi/2)$, where $R_{\rm m}$ is the magnetospheric radius. In keeping with the papers \cite{payne2004burial, priymak2011quadrupole, rossetto2023magnetically}, we choose
\begin{equation}
    M(\psi) = \frac{M_{\rm a}}{2}\frac{1-e^{-b\psi/\psi_*}}{1 - e^{-b}},
    \label{eq:dmdpsi_pm04}
\end{equation}
where $M_{\rm a}$ is the accreted mass, $b=\psi_*/\psi_{\rm a}$, $\psi_* = \psi(R_*,\pi/2) = B_*{R_*}^2/2$ is the magnetic flux at the pole and $B_* = |\mathbf{B}(R_*,0)|$ is the star's polar magnetic field. Note that, considering a dipolar magnetic field $\psi_{\rm d}(r,\theta)$,
\begin{equation}
    \label{psi_d}
    \psi_d(r,\theta) = \frac{\psi_*R_*}{r}\sin^2\theta, 
\end{equation}
we have a geometric interpretation of the parameter $b$. That is, $b$ is the ratio between the magnetosphere radius and the star radius $b=R_{\rm m}/R_*$. Furthermore, at the stellar surface, the half-opening angle of the polar cap is given by $\arcsin{(\sqrt{b})}$. The value of $b$ can be ascertained observationally by analysis of the X-ray spectra of white dwarfs; typical values are of the order $\sim10$ \cite{suleimanov2019hard,vermette2023constraining}.

\subsection{Boundary Conditions}
\label{sub-sec:boundary-conditions}

Equation \eqref{iso_grad_shafranov} -- the Grad-Shafranov equation -- is an elliptic partial differential equation and, as such, requires boundary conditions. We assume the system to be axisymmetric and posses north-south symmetry, therefore, it suffices to solve the equation in a quadrant of the plane, and we set: $\psi(r,0) = 0$, $\psi(R_*,\theta) = \psi_{\rm d}(R_*,\theta)$ and $\dd{\psi}/\dd{\theta} (r,\pi/2) = 0$. For the outer boundary condition, we impose that the field remains dipolar, as it was done in \citeauthor{rossetto2023magnetically}, \citeyear{rossetto2023magnetically} and \citeauthor{brunet2026relaxation}, \citeyear{brunet2026relaxation} \cite{rossetto2023magnetically, brunet2026relaxation}. That is, the magnetic dipole moment $m_{\rm d}$, defined by
\begin{equation}
    \label{mag_dip}
    m_{\rm d} = \frac{3r}{2}\int_{-1}^{1} \psi(r,\theta) \dd(\cos\theta),
\end{equation}
should not change with $r$ at the outer radius of the simulation $r_{\rm max}$. This implies the following Robin boundary condition:
\begin{equation}
    \label{robin_bc}
    \pdv{\psi}{r} + \frac{\psi}{r} = 0.
\end{equation}
Alternatively, \citeauthor{yeole2025investigating}, \citeyear{yeole2025investigating} have used current-free boundary conditions at the outer radius, by matching the numerical solution to the analytical small-$M_{\rm a}$ solution of PM04. Further considerations about this choice of boundary condition is discussed in Appendix \ref{sec:app_bc}.

As it is described in Section \ref{sec:numerical_methods}, we solve equations \eqref{iso_grad_shafranov} and \eqref{Fpsi} using an iterative scheme. For such, we need an ``initial condition'' -- a starting guess -- for the problem, and for that we use the dipole function \eqref{psi_d}. The iterative scheme mimics a time evolution of the system \cite{mouschovias1974static}, and, in this sense, we follow the deformations of an initial dipole to a final equilibrium state of the star's magnetic field and the accreted mass.

%% file: Sections/3-numerical_methods.tex
The numerical methods used in the present paper have been adapted from the neutron star literature \cite{payne2004burial, priymak2011quadrupole, rossetto2023magnetically}, but with a few improvements made. We explain the complete numerical method and its improvements in this section. In Section \ref{sub-sec:num_eqs}, we discuss the grid and the dimensionless equations; in Section \ref{sub-sec:iter_method}, we discuss the iterative self-consistent solution method; and, in Section \ref{sub-sec:convergence}, we discuss the convergence tests for the numerical solution.

\subsection{Dimensionless Equations and Grid}
\label{sub-sec:num_eqs}

In order to make equations \eqref{iso_grad_shafranov} and  \eqref{Fpsi} suitable for numerical integration, we introduce the grid coordinates
\begin{align}
    x &= \frac{r-R_*}{x_0},
    \\
    \mu &= \cos\theta,
\end{align}
where $x_0 = {c_s}^2{R_*}^2/(GM_*)$ is the pressure (and density) scale height. With these coordinates, equations \eqref{iso_grad_shafranov} and \eqref{Fpsi} become
\begin{align}
    \tilde{\Delta}^\ast\tilde{\psi} &= -Q_0\tilde{F}^{\prime}(\tilde{\psi})e^{-x},
    \label{GS_num}
    \\
    \tilde{F}(\tilde{\psi}) &= \frac{d\tilde{M}}{d\tilde{\psi}}\left[\int_{\mathcal{C}}d\tilde{s}(x+a)(1-\mu^2)^{1/2}|\tilde{\nabla}\Tilde{\psi}|^{-1}e^{-x}\right]^{-1},
    \label{Fpsi_num}
\end{align}
where the tilde represents dimensionless quantities, defined by $\tilde{\psi} = \psi/\psi_*$, $Q_0 = 4\pi {x_0}^4$, $\tilde{F} = F/F_0$, $F_0 = M_{\rm a}{c_{\rm s}}^2/{x_0}^3$, $\tilde{M} = M/M_{\rm a}$, $\tilde{s} = s/x_0$, $a = R_*/x_0$ and the dimensionless Grad-Shafranov operator $\tilde{\Delta}^\ast$ is
\begin{equation}
    \tilde{\Delta}^\ast = \frac{1}{(x + a)^2(1-\mu^2)}\left[\frac{\partial^2}{\partial x^2} + \frac{1-\mu^2}{(x + a)^2}\frac{\partial^2}{\partial\mu^2}\right].
    \label{eq:GS_operator_dimensionless}
\end{equation}
We use a $(N_x,N_\mu)$ grid of points, typically $(256,256)$. The grid is uniform in $\mu$, but non-uniform on $x$. On $x$ we make the variable change,
\begin{equation}
    \label{x_tilde}
    \tilde{x} = \log(x+1),
\end{equation}
and then uniformly distribute the grid points in $\tilde{x}$. This is done in order to have several grid points near the surface and capture the exponential decay of both the pressure and the density.

\subsection{Iterative Self-Consistent Method}
\label{sub-sec:iter_method}

\begin{figure}
    \centering
    \includegraphics[width=
    \linewidth]{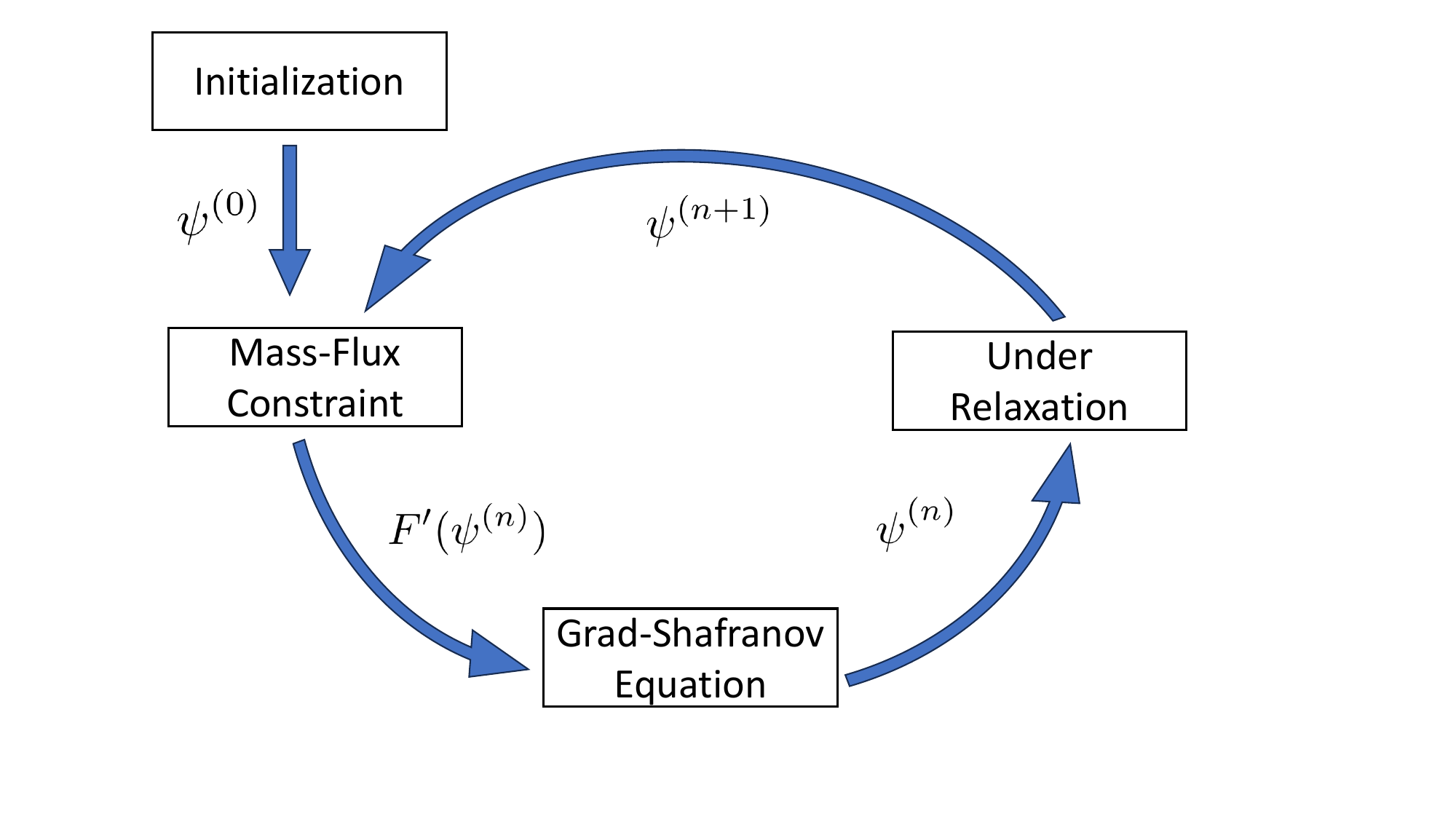}
    \caption{Diagram of the iterative scheme for a self-consistent solution of equations \eqref{GS_num} and \eqref{Fpsi_num}.}
    \label{fig:num_method}
\end{figure}

To solve equations \eqref{GS_num} and \eqref{Fpsi_num}, we utilize an iterative self-consistent method first described by \citeauthor{mouschovias1974static} \cite{mouschovias1974static} and adapted for the neutron star mountain problem by \citeauthor{payne2004burial} \cite{payne2004burial}. The method consists in starting with an initial guess $\psi^{(0)}$ for the flux function, using this function to calculate the source term \eqref{Fpsi_num}, solving the Grad-Shafranov equation \eqref{GS_num} with the calculated source term, under-relaxing the solution, calculate a new source term with the under-relaxed solution, solve the equation with the new source term, and repeat this procedure until a certain convergence criterion is met. Figure \ref{fig:num_method} depicts this solution method.

The initial function $\psi^{(0)}$ is normally taken to be a dipole,
\begin{equation}
    \label{init_dipole}
    \psi_d(r,\theta) = \frac{\psi_* R_*\sin^2\theta}{r},
\end{equation}
but it can also be chosen to be the numerical solution of a previous simulation with a lower value of accreted mass $M_{\rm a}$. Then, we compute the source term \eqref{Fpsi_num} using the \texttt{Python} package \texttt{Contourpy}. We use the package's functionality to find $N_{c}=N_x-1$ \footnote{This is the optimal choice for the number of contours for the reduction of the numerical residuals. For a discussion on this, see \cite{payne2004burial}.} isocontours $\mathcal{C}$ of $\psi$ and we use bilinear interpolation to compute the relevant field values on $\mathcal{C}$. Then, we use Simpson's method to compute the integral for every contour line $\mathcal{C}$, hence obtaining $F(\psi)$. Due to numerical fluctuations, finite difference gives a poor approximation for the derivative of $F(\psi)$. To suppress numerical noise, we fit an 8th-degree polynomial and then take its analytical derivative for the value of $F'(\psi)$. Finally, $F'(\psi)$ is mapped back to the grid by bilinear interpolation.

After the procedures explained previously, one has to solve equation \eqref{GS_num}, which is defined in the whole grid. For that, we take centred finite differences and obtain
\begin{equation}
    \label{finite_diff}
    a_{i,j}\tilde{\psi}_{i+1,j} +
    b_{i,j}\tilde{\psi}_{i-1,j} +
    c_{i,j}\tilde{\psi}_{i,j+1} +
    d_{i,j}\tilde{\psi}_{i,j-1} +
    e_{i,j}\tilde{\psi}_{i,j} =
    f_{i,j}
\end{equation}
where $i$ and $j$ are the indexes that label the $(x,\mu)$ grid, $\tilde{\psi}_{i,j} = \tilde{\psi}(x_i,\mu_j)$, and $a$, $b$, $c$, $d$, $e$ and $f$ are the matrix coefficients that appear after the finite difference of equation \eqref{GS_num}. The boundary conditions (see Section \ref{sub-sec:boundary-conditions}) are encoded in the boundary values of the matrix coefficients. We `flatten' the equation using a master index $k = i + N_x j$, thus obtaining a set of linear equations of the traditional form $A\cdot\mathbf{X}=\mathbf{b}$. From the properties of the finite difference scheme \eqref{finite_diff}, the matrix $A$ is sparse, containing only five non-empty diagonals. Therefore, the use of sparse solvers is appropriate, and we solve the system of equations directly using \texttt{UMFPACK} \cite{davis2004algorithm} as implemented in \texttt{Python}'s \texttt{Scipy} library \cite{virtanen2020scipy}. This solution method for the Grad-Shafranov equation is different from the commonly applied successive over-relaxation (SOR) in the neutron star literature \cite{payne2004burial, priymak2011quadrupole, rossetto2023magnetically, yeole2025investigating, brunet2026relaxation}. The sparse solver solves the system \eqref{finite_diff} with much greater precision than the SOR method -- as it is a direct solver -- and, in our simulations, the computational cost was comparable to that of SOR.

With the solution $\psi_{\rm GS}^{(n)}$ of the Grad-Shafranov equation \eqref{GS_num}, we under-relax the solution via
\begin{equation}
    \label{under_relax}
    \psi^{(n)} = \Theta^{(n)} \psi_{\rm GS}^{(n)} + \left(1-\Theta^{(n)}\right)\psi^{(n-1)},
\end{equation}
where $0<\Theta^{(n)}\leq 1$. This under-relaxation scheme aids convergence of the non-linear equation \eqref{GS_num} \cite{jardin2010computational}. We then use $\psi^{(n)}$ as the new initial guess, and we repeat the process explained above. Iterations continue until $\psi^{(n)}$ and $\psi^{(n-1)}$ are sufficiently close, quantified by their residuals defined in Section \ref{sub-sec:convergence}.

\subsection{Convergence Tests}
\label{sub-sec:convergence}

To verify the convergence of the physical system, we track the mean residual of $\psi$ across the grid, defined as
\begin{equation}
    \label{eq:residual}
    \left<\frac{\Delta\psi}{\psi}\right>^{(n)} = \frac{1}{N_x N_{\mu}}\sum_{i,j}\frac{\left|\psi_{i,j}^{(n)} - \psi_{i,j}^{(n-1)}\right|}{\left|\psi_{i,j}^{(n)}\right|}.
\end{equation}
We consider that the system has converged to a `final' equilibrium state if the mean residual is $10^{-5}$ or less. It is important to note that this residual is of the iterative self-consistent method and not of the sparse solver. The sparse solver obtains a solution near-machine precision; therefore, the self-consistent iterative scheme dominates the total residual.

Additionally, we verify the accuracy of the obtained solution $\psi$ by checking the mass in the simulation domain. Equation \eqref{rho} gives us the density of the mountain in the whole domain as a function of the solution $\psi$ and other grid quantities. We calculate the integral of $\rho$ in the whole domain using the Simpson's method, and we compare the result with half of $M_{\rm a}$, which is a user-defined quantity. We deem the solution suitable if the relative mass error is of the order of a few percent.

%% file: Sections/4-hydromag_structure.tex
In this section, we present the numerical results obtained by applying the methods of Section \ref{sec:numerical_methods}. In Section \ref{sub-sec:hydromag_structure}, we present the hydromagnetic structure of the magnetically confined mountain, focusing on the deformation of the magnetic field and the plasma's density. In Section \ref{sub-sec:max_ma}, we show the maximum mountain mass that the magnetic field can sustain for a range of the system's parameters. Unless it is otherwise stated, we adopt the following fiducial values in the simulations: $M_* = 0.8 M_\odot$, $R_* = 7\times10^8\,\mathrm{cm}$, $B_* = 5\times10^{7}\,\mathrm{G}$, $c_{\rm s} = 4.67\times10^{6}\,\mathrm{cm/s}$ ($T = 1.59\times 10^5\,\mathrm{K}$), $b=10$, $M_{\rm a} = 10^{-3}M_\odot$.

\subsection{Hydromagnetic Structure}
\label{sub-sec:hydromag_structure}

\begin{figure*}
    \centering
    \begin{subfigure}[b]{0.49\textwidth}
        \centering
        \includegraphics[height=0.35\textheight]{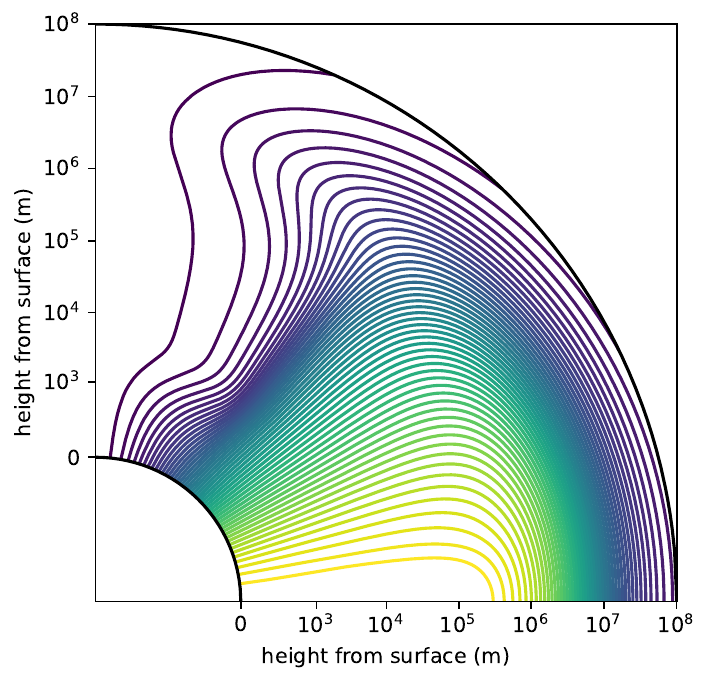}
        \caption{}
        \label{fig:psi_contours}
    \end{subfigure}
    \begin{subfigure}[b]{0.49\textwidth}
        \centering
        \includegraphics[height=0.35\textheight]{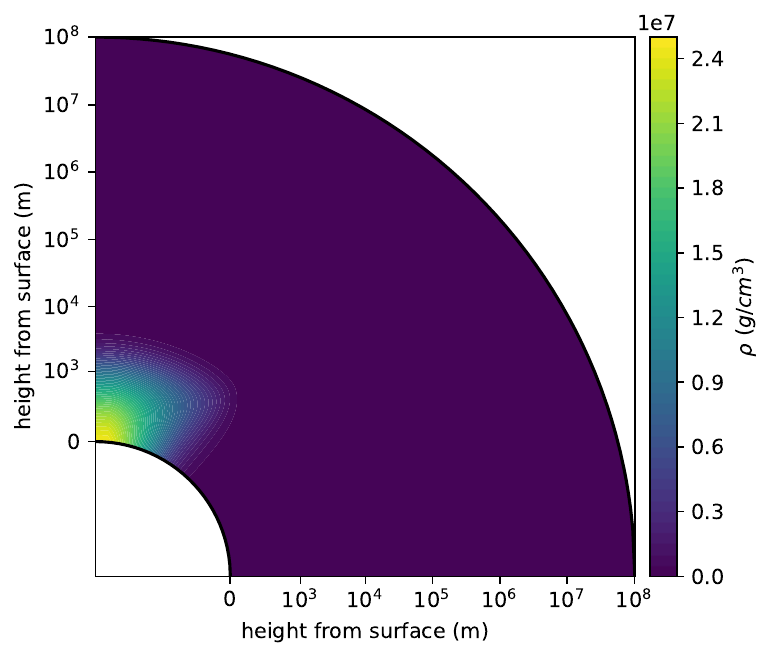}
        \caption{}
        \label{fig:density}
    \end{subfigure}
    \caption{Simulation of a magnetically confined mountain on a white dwarf's magnetic pole, the systems parameters used were $M_* = 0.8 M_\odot$, $R_* = 7\times10^8\,\mathrm{cm}$, $c_{\rm s} = 4.67\times10^{6}\,\mathrm{cm/s}$, $B_* = 5\times10^{7}\,\mathrm{G}$, $M_{\rm a} = 10^{-3}M_\odot$. (a) Magnetic field lines (isocontours of $\psi$). (b) Mountain's density.}
    \label{fig:model_simulation}
\end{figure*}

Figure \ref{fig:model_simulation} shows the hydromagnetic structure of an accreted mountain of mass $10^{-3}M_\odot$ in the polar region of a white dwarf. In both plots, a logarithmic scale was used for the height from the surface\footnote{The logarithmically scaling $\log_{10}(r-R_*(1-100/R_*))$ was used instead of the simple $\log_{10}(r-R_*)$ as to avoid a coordinate singularity at the star's surface.} to better visualise the mountain's effects. Panel \ref{fig:psi_contours} shows the contours of the function $\psi$, which, as a consequence of Equation \eqref{mag_field}, coincides with the magnetic field lines. It is visible that the mountain deforms the initial dipolar field. Panel \ref{fig:density} shows the density of the accreted mountain. The mountain spreads equatorwards, but remains magnetically confined. The maximum density is $2.48\times10^{7}\,\mathrm{g/cm^3}$ and it is obtained on the surface at the magnetic pole. This value is artificially high due to the simple isothermal equation of state and the solid surface condition. For the neutron star case, it is known that relaxing these two conditions, we arrive at lower values for the maximum density \cite{priymak2011quadrupole, wette2010sinking}. Near the magnetic pole, the mountain is purely sustained by hydrodynamical balance and, therefore, the density follows $\rho(r) = \rho_{\rm b} e^{-(r-R_*)/x_0}$, where $\rho_{\rm b}$ is the density at the base of the mountain and $x_0 \approx 1.01\times10^{5}\,\mathrm{cm}$ is the density scale-height. At higher colatitudes, the mountain is partially supported by the magnetic force and the simple exponential decay does not hold.

\begin{figure*}
    \centering
    \begin{subfigure}[b]{0.49\textwidth}
        \centering
        \includegraphics[width=\textwidth]{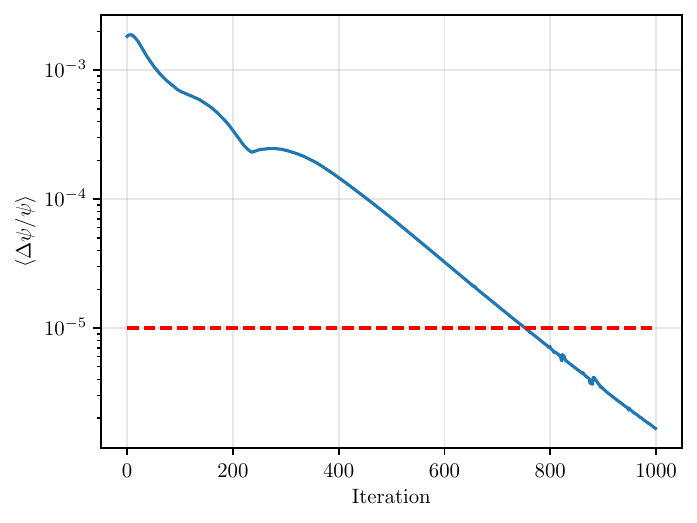}
        \caption{}
        \label{fig:psi_residuals}
    \end{subfigure}
    \begin{subfigure}[b]{0.49\textwidth}
        \centering
        \includegraphics[width=\textwidth]{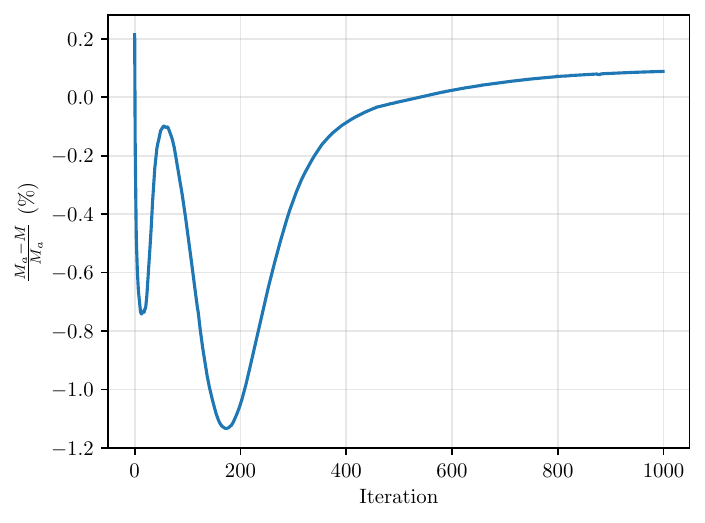}
        \caption{}
        \label{fig:mass error}
    \end{subfigure}
    \caption{Convergence tests for the simulations of Figure \ref{fig:model_simulation}. (a) $\psi$ residuals of the self-consistent iterative scheme. (b) Total mass error in the simulation domain.}
    \label{fig:convergence}
\end{figure*}

The convergence tests for the simulation presented in Figure \ref{fig:model_simulation} are done by the two methods explained in Section \ref{sub-sec:convergence}, and the results are presented in Figure \ref{fig:convergence}. Panel \ref{fig:psi_residuals} shows the residuals of the self-consistent iterative scheme, and it demonstrates that a final equilibrium is found with a residual smaller than $10^{-5}$. Panel \ref{fig:mass error} shows that the absolute mass error through the simulation domain stays mostly within $1\%$.

\subsection{Maximum Accreted Mass}
\label{sub-sec:max_ma}

The magnetic field deformations shown in \ref{fig:psi_contours} are dependent on the amount of accreted mass $M_{\rm a}$. For small $M_{\rm a}$, the magnetic tension is much greater than the pressure and the gravitational force on the mountain, which makes the mountain accumulate mass without affecting the dipolar structure of the magnetic field. For sufficiently big $M_{\rm a}$, the gravitational force makes the mountain spread equatorwards, dragging with it the frozen-in magnetic field lines. The distorted equilibrium is a result of the balance between pressure, gravitational and magnetic forces, as it is described by equation \ref{grad_shafranov} -- the Grad-Shafranov equation. From our numerical simulations, we observe that after a certain value of $M_{\rm a}$, the solution does not converge and both the $\psi$ residual \eqref{eq:residual} and the mass error grow beyond our defined convergence thresholds. Near this critical $M_{\rm a}$ value, the magnetic field lines become highly deformed and start to reconnect, forming magnetic islands. These features are responsible for the observed loss of equilibrium and can only be properly treated in full magnetohydrodynamic simulations. Highly distorted magnetic field lines are also subject to a plethora of MHD instabilities, such as Rayleigh–Taylor, sausage and kink instabilities,
all of which are not in the scope of this paper. This behaviour is also observed in the neutron star literature, for a variety of numerical approaches \cite{payne2004burial, priymak2011quadrupole, mukherjee2017revisiting, fujisawa2022magnetically, rossetto2023magnetically, yeole2025investigating}.

To pinpoint the maximum accreted mass $M_{\rm max}$ that is allowed by the Grad-Shafranov equilibrium, we employ the binary search algorithm, for which we give a low mass value -- where we know the solution converges -- and a high mass value -- where we know the solution does not converge. Typically, we use the value $10^{-9} M_\odot$ for the lower bound and $10^{-1} M_\odot$ for the upper bound. If at some point during the self-consistent iteration scheme, the mass error is equal to or greater than $10\%$, we halt the simulation and deem it not converged. We stop the binary search when we know $M_{\rm max}$ value within $1\%$.  We use this search algorithm on the $(c_{\rm s}, B_*)$ parameter space. For the magnetic field,  we consider the values in the range $1.00 \times 10^5\, \mathrm{G} \leq B_*\leq 3.00 \times 10^8\, \mathrm{G}$, as compatible with magnetic white dwarfs \cite{dufour2017montreal, 2023ApJ...944...56A, 2025A&A...698A.106S}. For the speed of sound $c_{\rm s}$, we relate it with the mountain's temperature using the isothermal perfect fluid equation,  
\begin{equation}
    \label{cs}
    c_s = \sqrt{\frac{k_BT}{\mu m_H}},
\end{equation}
where $k_B$ is Boltzmann's constant, $T$ is the temperature, $\mu$ is the mean molecular weight and $m_H$ is the mass of hydrogen. We consider $\mu=2$ and the temperature to vary in the range $2.00 \times 10^{4}\,\mathrm{K}\leq T\leq 1.00 \times 10^{7}\,\mathrm{K}$. The lower range of the temperature is around the temperature of the surface of white dwarfs \cite{dufour2017montreal, 2023ApJ...944...56A, 2017MNRAS.466.2855P}, while the upper range is the temperature associated with the heated accreted mass by the shock in the accretion column \cite{2005MNRAS.360.1091S, jose2025hydrodynamic, 2010A&A...520A..25Y}. Realistically, the mountain will have a temperature gradient, but this is not handled in our isothermal treatment. 

\begin{figure*}
    \centering
    \begin{subfigure}[b]{0.49\textwidth}
        \centering
        \includegraphics[width=\textwidth]{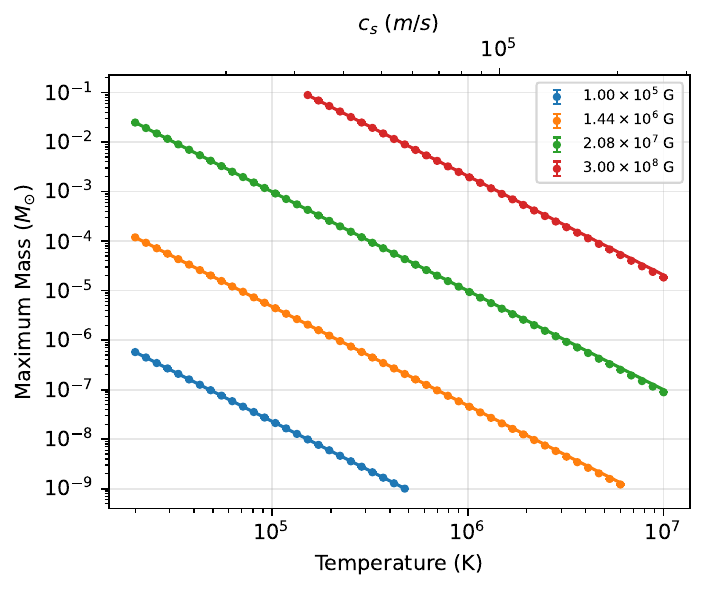}
        \caption{}
        \label{fig:max_mass_T}
    \end{subfigure}
    \begin{subfigure}[b]{0.49\textwidth}
        \centering
        \includegraphics[width=\textwidth]{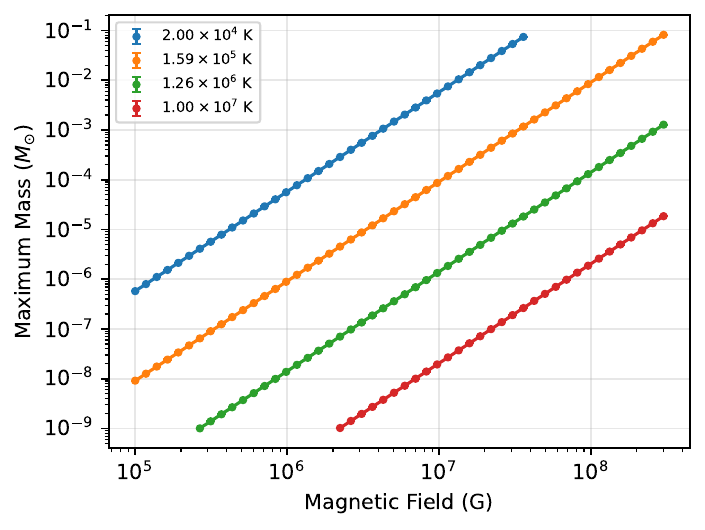}
        \caption{}
        \label{fig:max_mass_B}
    \end{subfigure}
    \caption{Maximum accreted mass as a function of the system's parameters. (a) Maximum accreted mass as a function of the mountain's temperature, for selected values of the magnetic field. (b) Maximum accreted mass as a function of the star's magnetic field, for selected values of temperature.}
    \label{fig:max_ma}
\end{figure*}

The results of our maximum mass search are displayed in Figure \ref{fig:max_ma}. Panels \ref{fig:max_mass_T} and \ref{fig:max_mass_B} show the dependence of the maximum $M_{\rm a}$ on $c_{\rm s}$ and $B_*$, respectively. For each parameter, there is a clear power-law relationship between the variables. A fit to the points shows that the power-law exponent for the magnetic field is $2$ and for the temperature is $-2$ (equivalently, $-4$ for the sound speed). Furthermore, the power-law coefficient is the same for both the magnetic field and temperature dependencies, allowing us to write
\begin{equation}
    \label{max_mass_formula_fit}
    M_{\rm max} = 4.07\times10^{-3}\,M_\odot
    \left(\frac{T}{10^5 \, \mathrm{K}}\right)^{-2}\left(\frac{B_*}{5\times10^7 \, \mathrm{G}}\right)^{2}.
\end{equation}
This power-law dependency is encapsulated in PM04's formula for the critical mass $M_{\rm c}$ given by
\begin{equation}
    \label{Mc_approximation}
    M_{\rm c} = \frac{GM_*B_*^2 {R_*}^2}{8{c_{\rm s}}^4}.
\end{equation}
In PM04, equation \eqref{Mc_approximation} was derived from the analytical solution of the Grad-Shafranov problem for the small-$M_{\rm a}$ limit. Making a fit of the type $aM_{\rm c}$, give us $a = (3.2\pm0.1)\times10^{-2}$, that is, the maximum confined mountain mass is $3\%$ of the critical mass $M_{\rm c}$.

%% file: Sections/5-mag_burial.tex
Accreted mountains on white dwarfs deform the magnetic field lines, see Figure \ref{fig:psi_contours}, and change the topology of the field. This process is called magnetic burial. In Section \ref{sub-sec:mag_dip}, we discuss the screening of the star's magnetic field and the reduction of the magnetic dipole moment; and, in Section \ref{sub-sec:mag_multipoles}, we compute the excitation of the magnetic multipole moments and compute their relative strengths.

\subsection{Magnetic Dipole Moment}
\label{sub-sec:mag_dip}

\begin{figure*}
    \centering
    \begin{subfigure}[b]{0.49\textwidth}
        \centering
        \includegraphics[width=\textwidth]{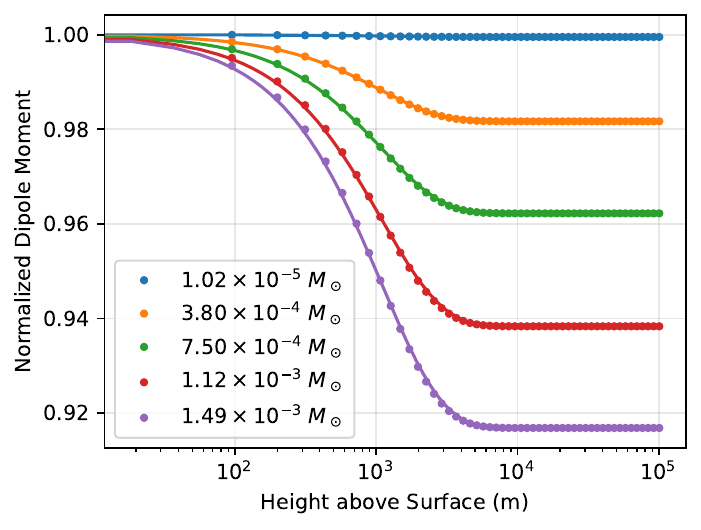}
        \caption{}
        \label{fig:dip_mom_per_r}
    \end{subfigure}
    \begin{subfigure}[b]{0.49\textwidth}
        \centering
        \includegraphics[width=\textwidth]{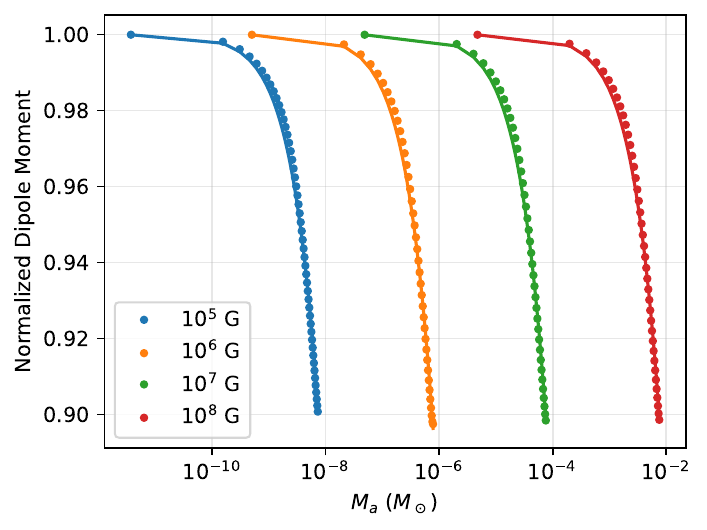}
        \caption{}
        \label{fig:dip_mom_per_Ma}
    \end{subfigure}
    \caption{Magnetic dipole moment for several $M_{\rm a}$. (a) Cumulative dipole moment per height above the surface for five different values of $M_{\rm a}$. (b) Integrated dipole moment per accreted mass $M_{\rm a }$ for different values of magnetic field $B_*$.}
    \label{fig:dip_mom}
\end{figure*}

The form of the magnetic field, given by equation \eqref{mag_field}, allows us to write the magnetic multipole moments $m_{l}$ at distance $r$ as
\begin{equation}
    \label{mag_multipoles}
    m_{l}(r) = \frac{l(2l+1)}{2(l+1)} r^l \int_{-1}^{1} \psi(r,\mu) \dv{P_l(\mu)}{\mu}\dd{\mu},
\end{equation}
where $P_l$ is the Legendre polynomial of order $l$. From the assumed north-south symmetry of $\psi(r,\mu)$ and the parity of Legendre polynomials, all the even multipoles vanish. Figure \ref{fig:dip_mom} shows the dipole moment ($l=1$) of white dwarfs after the accretion of polar mountains. Panel \ref{fig:dip_mom_per_r} shows the dipole moment, normalised by its surface value, per height and for different values of accreted mass. To facilitate visualisation, the plot was done for every fifth spatial point. We notice the effect of screening currents in the mountain that partially bury the star's magnetic dipole moment; then, outside a few scale heights, the value asymptotes to a constant. The higher the accretion, the bigger the screening of the star's dipole moment. The line that joins the points is an exponential fit of the form
\begin{equation}
    \label{m_per_r_fit}
    m_1(r) = (1-m_\infty)e^{-(r-R_*)/\lambda}+m_\infty.
\end{equation}
For all curves, we find a consistent value of $\lambda = (1.07 \pm 0.3)\times 10^{5}\,\mathrm{cm}$ which is compatible with the scale height of the mountain, $x_0 = 1.01\times 10^{5}\,\mathrm{cm}$. For $m_\infty$, which represents the asymptotic value of the normalised magnetic dipole moment, we obtained $0.999$, $0.981$, $0.962$, $0.938$ and $0.916$ for $M_{\rm a}/M_\odot$ equal to $1.02\times10^{-5}$, $3.80\times10^{-4}$, $7.50\times10^{-4}$, $1.12\times10^{-3}$ and $1.49\times10^{-3}$, respectively.

On Panel \ref{fig:dip_mom_per_Ma}, we plot the value of the dipole moment at the outer boundary of the simulation, $r_{\rm max} = 100 x_0$, normalised by the value at the surface, per accreted mass for different values of the star's pre-accretion magnetic field. Again, we see a decrease in the normalised magnetic dipole moment with increasing accreted mass. We also see that higher magnetic fields can confine higher accreted masses. The simulations were run near the system's maximum mass, as per Figure \ref{fig:max_mass_B}, and we notice that the maximum relative screening of the dipole moment is independent of the magnetic field and its value is around $10\%$. We fit the data using the expression
\begin{equation}
    \label{m_analytical}
    \frac{m_1(r_{\rm max})}{m_1(R_*)} = \left(1+\frac{M_{\rm a}}{M_{\rm c}}\right)^{-1},
\end{equation}
where $M_{\rm c}$ is the characteristic mass parameter. The fitted curves are shown as the lines connecting the points in \ref{fig:dip_mom_per_Ma}. The parameter $M_{\rm c}$ obtained from the fit (labelled ``fit'') is shown in Table \ref{tab:Mc}, together with the value calculated by the analytical approximation (labelled ``an'') given by equation \eqref{Mc_approximation}. We notice that the fitted values are near the values obtained by the approximation; nevertheless, they are statistically different. This discrepancy is expected since we have calculated the full numerical solution of equations \eqref{iso_grad_shafranov} and \eqref{Fpsi}, while the analytical approximation uses an explicit form of $F(\psi)$ that is appropriate for small-$M_{\rm a}$ (rf. \cite{payne2004burial}).

\begin{table}
\centering
\begin{tabular}{|c|c|c|}
\hline
$B_*$              & $M_{\rm c, fit}\, (M_\odot)$   & $M_{\rm c, an}\, (M_\odot)$ \\ \hline
$10^5\,\mathrm{G}$ & $(6.75 \pm 0.04) \times 10^{-8}$ & $6.88 \times 10^{-8}$       \\ \hline
$10^6\,\mathrm{G}$ & $(6.74 \pm 0.04) \times 10^{-6}$ & $6.88 \times 10^{-6}$       \\ \hline
$10^7\,\mathrm{G}$ & $(6.74 \pm 0.05) \times 10^{-4}$ & $6.88 \times 10^{-4}$       \\ \hline
$10^8\,\mathrm{G}$ & $(6.74 \pm 0.05) \times 10^{-2}$ & $6.88 \times 10^{-2}$       \\ \hline
\end{tabular}
\caption{Critical mass values $M_{\rm c}$ obtained by the curve fit and by the analytical expression \eqref{Mc_approximation}.}
\label{tab:Mc}
\end{table}

\subsection{Higher Magnetic Multipole Moments}
\label{sub-sec:mag_multipoles}

\begin{figure*}
    \centering
    \begin{subfigure}[b]{0.49\textwidth}
        \centering
        \includegraphics[width=\textwidth]{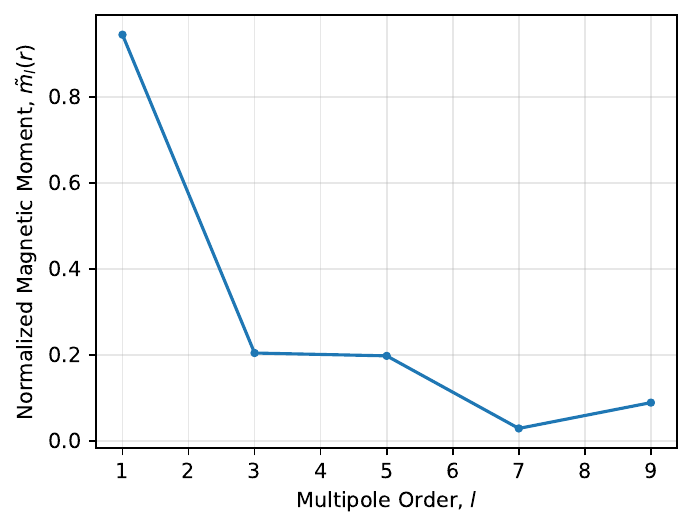}
        \caption{}
        \label{fig:multipoles_fraction}
    \end{subfigure}
    \begin{subfigure}[b]{0.49\textwidth}
        \centering
        \includegraphics[width=\textwidth]{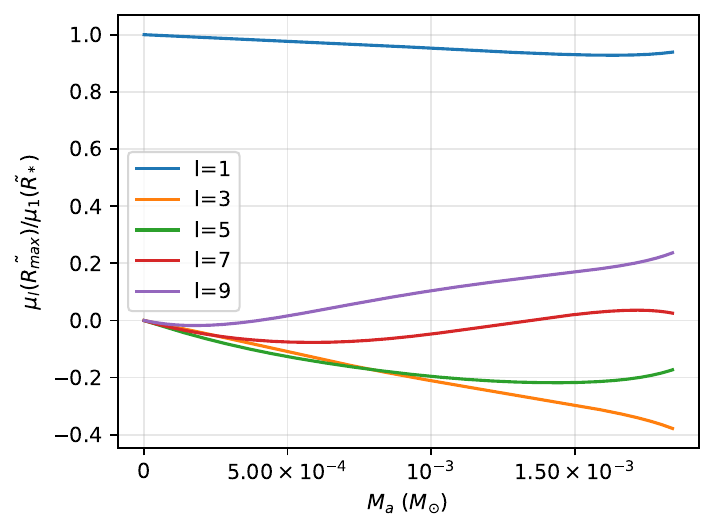}
        \caption{}
        \label{fig:multipoles_per_Ma}
    \end{subfigure}
    \caption{Magnetic multipole moments. (a) Fractional intensity of each multipole of order $l$ for the magnetic field configuration displayed in Figure \ref{fig:model_simulation}. (b) Multipole moments per accreted mass $M_{\rm a}$, the system's parameters are the same as for Figure \ref{fig:model_simulation} but with varying $M_{\rm a}$.}
\end{figure*}

To compare different multipole moments, we use the normalised multipoles $\tilde{m}_{l}$ \cite{suvorov2020recycled, fujisawa2022magnetically, yeole2025investigating}, defined by
\begin{equation}
    \label{norm_multipoles}
    \tilde{m}_{l}(r) = \frac{2(l+1)}{l(2l+1)} \frac{1}{r^{l-1}} \frac{m_{l}(r)}{m_{1}(R_*)}.
\end{equation}
Panel \ref{fig:multipoles_fraction} shows the post-accretion multipolar fraction of the magnetic field of the star for the simulation displayed in \ref{fig:model_simulation}. It is still dominantly dipolar, but with a non-negligible contribution from higher moments. Panel \ref{fig:multipoles_per_Ma} depicts the dependency of the multipoles on the accreted mass $M_{\rm a}$. For low values of $M_{\rm a}$, the accreted mass does not significantly distort the magnetic field, and the dipole moment is the only one present. As $M_{\rm a}$ increases, the dipole decreases, and the magnitude of the other multipoles increases.

%% file: Sections/6-conclusion.tex
In this work, we applied the theory of self-consistent magnetically confined mountains to the astrophysical problem of accreting magnetic white dwarfs. We have considered the accreted mass to be in magnetohydrostatic equilibrium, and we have analysed the conditions under which this equilibrium holds. Previous works on magnetic accretion onto white dwarfs have considered simpler spherical or cylindrical models, with order-of-magnitude estimates of the maximum accreted mass based on broad assumptions about magnetic confinement \cite{cumming2002magnetic,scaringi2022triggering,hameury1985magnetohydrostatics,zhang2006bottom,zhang2009there}. In this paper, we solve the magnetostatic fluid equations, i.e., the Grad-Shafranov equation \eqref{iso_grad_shafranov}, together with the mass-flux constraint \eqref{Fpsi}. 

In our investigations, we consider both an individual, representative white dwarf and a batch of simulations for different system parameters. For a fiducial star with $M_* = 0.8 M_\odot$, $R_* = 7\times10^8 \, \mathrm{cm}$, $B_* = 5\times10^{7}\,\mathrm{G}$ and $T = 1.59\times10^{5} \, \mathrm{K}$  (see Figure \ref{fig:model_simulation}), we found that $M_{\rm a} \sim 10^{-3}M_\odot$ is needed for the magnetic field lines to be reasonably distorted away from the dipolar configuration. For this case, the mountain's scale height is $x_0\approx 1.01 \times 10^5\, \mathrm{cm}$, and this is the characteristic length for the polar density and pressure decays -- at higher colatitudes, the magnetic forces also support the mountain. The accreted matter reduces the star's dipole moment by $\sim5 \%$, this effect is known as magnetic field burial. The screening currents are limited near the surface, making the dipole moment reach its far-source value within a few scale heights (see Figure \ref{fig:dip_mom_per_r}). The final configuration of the magnetic field is still dominantly dipolar, but with excited higher (odd) multipole moments (see Figure \ref{fig:multipoles_fraction}). As the accreted mass $M_{\rm a}$ increases, the magnetic field lines get progressively more distorted until they start to reconnect, and, subsequently, the system loses equilibrium, and we do not find converging solutions. 

In our batch simulations, we searched for the maximum allowed $M_{\rm a}$ and how this value depends on the system's parameters.  We discovered that the maximum allowed mass $M_{\rm max}$ scales as
\begin{equation}
    \label{max_mass_formula}
    M_{\rm max} = 4.07\times10^{-3}\,M_\odot
    \left(\frac{T}{10^5 \, \mathrm{K}}\right)^{-2}\left(\frac{B_*}{5\times10^7 \, \mathrm{G}}\right)^{2}.
\end{equation}   
\noindent The functional dependencies of the equation \eqref{max_mass_formula} are the same as those for the critical mass $M_{\rm c}$ in the small-accretion limit calculated by  \citeauthor{payne2004burial}, \citeyear{payne2004burial} \cite{payne2004burial}, but with a different leading factor since we computed the cited relation using the non-approximate, self-consistent solution to the Grad-Shafranov equation.

Another important result from our simulations is regarding magnetic field burial. We have calculated the effect of the accreted mass $M_{\rm a}$ on magnetic dipole moment reduction and on the excitation of multipolar modes. As expected, larger $M_{\rm a}$ produces greater attenuation of the white dwarf's dipole moment. The obtained relationship between the normalised (by the pre-accretion value) magnetic dipole moment and $M_{\rm a}$ is
\begin{equation}
    \label{m_fit}
    \frac{m_1(r_{\rm max})}{m_1(R_*)} = \left(1+\frac{M_{\rm a}}{M_{\rm c}}\right)^{-1},
\end{equation}
where $r_{\rm max}$ being the outer radius of the simulation and where $M_{\rm c}$ given in Table \ref{tab:Mc}, and, approximately, by equation \eqref{Mc_approximation}. This result is, however, limited by the maximum mass $M_{\rm max}$ that the system can confine, given by equation \eqref{max_mass_formula}. We have verified that when $M_{\rm a}$ is near the maximum, magnetic burial reduces the magnetic field by $\sim 10\%$ (see Figure \ref{fig:dip_mom_per_Ma}), independently of the pre-accretion magnetic field value. We also observe an increase in the higher multipole moments associated with the decrease of the dipole moment; only odd multipoles are excited given the assumed north-south symmetry. 

As our work is a first pass on the full MHD model of the problem, we have considered some simplifying assumptions that could be relaxed in future work. These include:

(i) Isothermal equation of state. For simplicity of equations and ease of numerical implementation, we have considered an isothermal equation of state, $p={c_{\rm s}}^2\rho$, for the magnetically confined mountain on the white dwarf. The theory of Section \ref{sec:hydromag_structure} and the methods of Section \ref{sec:numerical_methods} can be adapted to more general equations of state, as was done in the neutron star literature \cite{priymak2011quadrupole}.

(ii) No mountain sinking. We have considered a dipolar boundary condition for the magnetic field at the white dwarf's surface, and also assumed that the mountain does not sink into the star. In the neutron star literature, this assumption has been relaxed by considering an atmospheric layer and the mixing of mountain matter with the star's matter \cite{wette2010sinking,yeole2025investigating, brunet2026relaxation}, these approaches are beyond the scope of this paper. 

(iii) Ideal MHD. We considered the accreted mountain to be a perfectly conducting plasma, such that ideal MHD holds. From this assumption, we have the integral condition \eqref{Fpsi} -- which is a fundamental piece of the self-consistent iterative scheme (see Section \ref{sec:numerical_methods}). The numerical results we obtained show that white dwarfs can easily accumulate high values of accreted mass (compared to their own mass), depending on the star's temperature and magnetic field. This means that magnetic and pressure gradients are relatively smooth and, therefore, ideal MHD is likely a reasonable approximation. Nevertheless, relaxation of the ideal MHD condition can be obtained by both semi-analytical and full MHD simulations \cite{vigelius2009resistive,suvorov2019relaxation,kulsrud2020anomalous,brunet2026relaxation}.

(iv) No rotation effects. Our model considers the total mass of the magnetically confined mountain without explicitly simulating the accretion phase that formed it. However, the physical accumulation of this mass could be limited by rotational mass-loss mechanisms, particularly in asynchronous cataclysmic variables such as intermediate polars, \cite{patterson1994, 2003MNRAS.338.1067P, Pelisoli2022}. In these systems, the propeller regime is activated when the magnetospheric radius exceeds the co-rotation radius ($R_{\rm m} > R_{\rm co}$) \cite{1995ApJ...449L.153W, 1999ApJ...520..276M}. Under these conditions, the white dwarf's magnetic field lines rotate faster than the local Keplerian velocity of the infalling material, establishing a centrifugal barrier. This interaction transfers angular momentum to the plasma, centrifugally ejecting it from the system and inhibiting further accretion \cite{becerra2018, 2022ApJ...941...28S}. Consequently, the interplay between the mass accretion rate, magnetic field strength, and stellar spin period might physically restrict the formation of the massive mountains we explore here.

(v) No dynamical effects. The Grad-Shafranov equation used in this work describes the static equilibrium of the polar accreted mass onto white dwarfs. Therefore, the simulations presented here do not encode any dynamical effects, including details of the accretion process and the development of hydrodynamic and MHD oscillations and instabilities. These effects are crucial to evaluate the stability of the equilibrium solutions presented. To study the dynamical evolution of the accreted plasma, one should use full MHD codes, potentially using the Grad-Shafranov equilibria calculated here as initial conditions.

Our present model for magnetically confined mountains on white dwarfs has applications to other astrophysical scenarios. The presence of matter on the magnetic poles gives the star a non-zero ellipticity that, combined with a misalignment between the magnetic and rotation axis, makes the star serve as a source for continuous gravitational waves \cite{2017MNRAS.467.4484F, 2019MNRAS.490.2692K, 2020MNRAS.492.5949S, sousa2024prospects}. The waves generated this way can potentially be detected by future space-based detectors, such as LISA and TianQin, whose operating frequency band is in the interval of $10^{-4}$~Hz $-$ $10^{-1}$~Hz, and BBO and DECIGO, covering a frequency range of $0.01$~Hz to $10$~Hz \cite{AMARO/2017, 2016CQGra..33c5010L, 2006CQGra..23.4887H, 2017PhRvD..95j9901Y, 2006CQGra..23S.125K}. Furthermore, accreted mountains have high pressure at their bases, which can lead to thermonuclear explosions, such as novae and micronovae, and our model can provide information on the amount of mass, magnetic field, and temperature that is necessary for such explosions to happen \cite{scaringi2022triggering}. Both these implications will be studied in detail in future papers.

The theoretical and numerical methods used in this paper originated from the neutron star literature, in particular in PM04 \cite{payne2004burial}, with improved outer boundary condition (as it was firstly proposed by \citeauthor{rossetto2023magnetically}, \citeyear{rossetto2023magnetically} \cite{rossetto2023magnetically}), the use of sparse solvers for the Grad-Shafranov equation, and the performance of a binary search for the maximum accreted mass. The suitability of these methods were initially unknown for white dwarfs, given the difference in scales, but the methods have proven well-suited to this case as well. Nevertheless, there are important differences between the solutions obtained here and the ones for neutron stars. The larger radii of white dwarfs and reduced surface gravity make all the gradients smoother resulting in more absolute accreted mass, compared to neutron stars. Furthermore, the magnetic burial process seems to be attenuated for white dwarfs, with maximum masses decreasing the star's dipole moment only by $10\%$. Therefore, caution is needed when carrying over results and conclusions from the neutron star literature to the white dwarf case.

%% file: Sections/acknowledgments.tex
This study was financed, in part, by the São Paulo Research Foundation (FAPESP), Brasil. Process Number 2024/21854-2. MFS is also grateful for the financial support of FAPESP (grants No. 2026/06733-0, No. 2025/05794-2 and No. 2021/01089-1). DFG also thanks FAPESP for grant 2021/02120-0, and Conselho Nacional de Desenvolvimento Científico e Tecnológico (CNPq) for grant 304574/2024-4. The computations were performed using the Hydra HPC Cluster at the Laboratory of Theoretical Astrophysics, University of São Paulo, supported by FAPESP grant 2022/03972-2. We thank Prof. Dr. Grzegorz Kowal for his technical support in the usage of the Hydra cluster.

%% file: Sections/appendix.tex
The Grad-Shafranov equation is an elliptical partial differential equation and, as such, its solutions are determined by boundary conditions (BCs). In this section, we focus on the effects of the outer BC on the solution obtained by our self-consistent iterative scheme discussed in \ref{sub-sec:iter_method}.

In the neutron star literature, where the Grad-Shafranov method is commonplace, several choices have been made for the outer boundary condition: (a) fixed boundary conditions \cite{mukherjee2012phasedependent, mukherjee2017revisiting}; (b) outflow boundary condition $\partial\psi/\partial r = 0$ \cite{melatos2005gravitational, priymak2011quadrupole};  (c) dipolar boundary conditions \cite{rossetto2023magnetically,brunet2026relaxation}, as used in this work and explained in Section \ref{sub-sec:boundary-conditions}; and (d) current-free boundary conditions \cite{yeole2025investigating}. Option (a) it keeps a fixed dipole at the outer boundary; nevertheless, the fixed value has to be set beforehand, and then the solution is forced to that value at the boundary. Option (b) allows the dipole moment to change from its surface (or otherwise set) value, but it implies unphysical radial magnetic fields at the outer boundary (see equation \eqref{mag_field}). \citeauthor{rossetto2023magnetically}, \citeyear{rossetto2023magnetically} \cite{rossetto2023magnetically} introduced option (c), which imposes that the magnetic dipole moment does not change as you approach the outer boundary, this method corrects the two drawbacks of options (a) and (b). The recently developed option (d), introduced by \citeauthor{yeole2025investigating}, \citeyear{yeole2025investigating} \cite{yeole2025investigating}, imposes that the outer boundary is current-free by matching the solution there to the small-$M_{\rm a}$ analytical solution developed by \citeauthor{payne2004burial}, \citeyear{payne2004burial} \cite{payne2004burial}. This choice makes the dipole moment more well-behaved and allows higher multipolar moments to form.

\begin{figure}
    \centering
    \includegraphics[width=
    \linewidth]{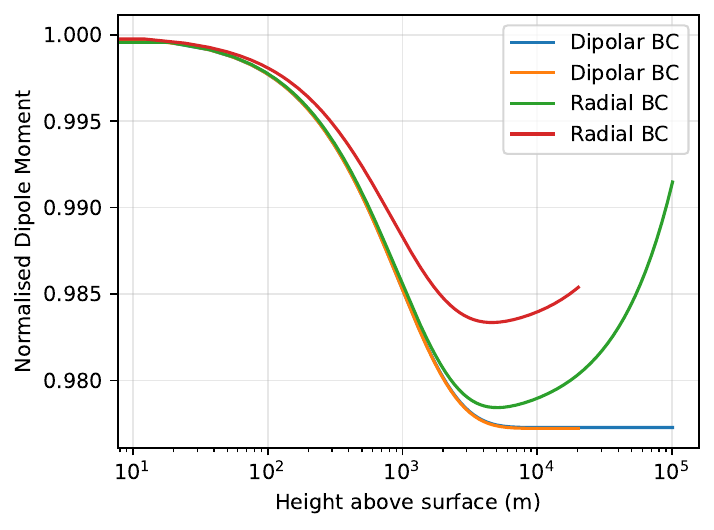}
    \caption{Dependence of the magnetic dipole moment on the choice of boundary condition at the outer radius. Green and red curves represent the radial magnetic field boundary condition, while blue and orange represent the dipolar choice.}
    \label{fig:dip_mom_BC}
\end{figure}

We compare the dipole boundary condition, explained in Section \ref{sub-sec:boundary-conditions}, with the outflow boundary condition for two values of $r_{\rm max}$, $100x_0$ and $20x_0$. The result is shown in Figure \ref{fig:dip_mom_BC}. For all boundary conditions, and for all $r_{\rm max}$, we observe that the normalised dipole moment has a unit value near the surface and decays with increasing radius. For the outflow boundary conditions, the dipole moment resurges as $r\to r_{\rm max}$, due to the unphysical radial magnetic field lines. On the other hand, the dipolar boundary condition decreases monotonically until it plateaus near $r_{\rm max}$. The latter behaviour is what is physically expected -- the mountain induces screening currents that partially shield the star's magnetic field, but a few scale heights above the mountain the dipole moment is expected to settle to a reduced value. We have also plotted the field lines, as in Figure \ref{fig:psi_contours}, but the resulting plots are very similar visually; the differences are more pronounced in the normalised dipole moment plot, Figure \ref{fig:dip_mom_BC}.

\begin{figure}
    \centering
    \includegraphics[width=
    \linewidth]{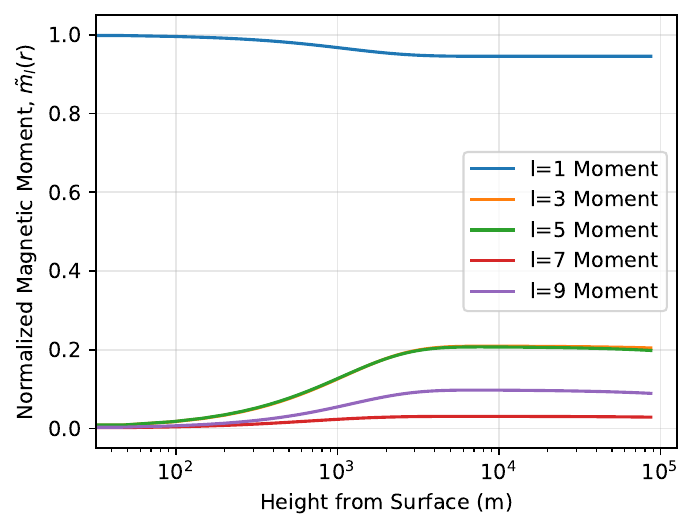}
    \caption{Radial dependency of the magnetic multipole moments for the magnetic field configuration displayed in Figure \ref{fig:psi_contours} for the dipolar boundary condition.}
    \label{fig:multipoles_per_r}
\end{figure}

\citeauthor{yeole2025investigating}, \citeyear{yeole2025investigating}, showed that the current-free boundary condition also makes the magnetic dipole moment decrease monotonically and plateaus near $r_{\rm max}$ \cite{yeole2025investigating}. Presumably, their approach also makes the higher multipoles plateau near the outer boundary. Figure \ref{fig:multipoles_per_r} shows the radial dependency of the multipole moments for our choice of boundary conditions, i.e., option (c). We notice that the multipoles with $l\geq3$ do approach an asymptotic value, albeit with an ever-so-slight decrease near $r_{\rm max}$. This shows that, even though our boundary condition was designed to keep the dipole moment unchanged near the outer boundary, we get the same behaviour for the higher multipoles. Therefore, we are justified in using the dipolar boundary condition as it gives us the expected physical result with a simple implementation.

%% file: references.bib
@article{brunet2026relaxation,
  title = {Relaxation of Magnetically-Confined Mountains on Accreting Neutron Stars through Cross-Field Mass Transport},
  author = {Brunet, Ryan and Melatos, Andrew and Rossetto, Pedro H B},
  date = {2026-04-24},
  journaltitle = {Monthly Notices of the Royal Astronomical Society},
  shortjournal = {Mon. Not. R. Astron. Soc.},
  volume = {548},
  number = {3},
  pages = {stag695},
  issn = {0035-8711, 1365-2966},
  doi = {10.1093/mnras/stag695}
}

@article{cumming2002magnetic,
  title = {Magnetic Field Evolution in Accreting White Dwarfs},
  author = {Cumming, A.},
  date = {2002-07-01},
  journaltitle = {Monthly Notices of the Royal Astronomical Society},
  shortjournal = {Mon. Not. R. Astron. Soc.},
  volume = {333},
  number = {3},
  pages = {589--602},
  issn = {0035-8711, 1365-2966},
  doi = {10.1046/j.1365-8711.2002.05434.x}
}

@article{davis2004algorithm,
  title = {Algorithm 832: {{UMFPACK V4}}.3---an Unsymmetric-Pattern Multifrontal Method},
  shorttitle = {Algorithm 832},
  author = {Davis, Timothy A.},
  date = {2004-06},
  journaltitle = {ACM Transactions on Mathematical Software},
  shortjournal = {ACM Trans. Math. Softw.},
  volume = {30},
  number = {2},
  pages = {196--199},
  issn = {0098-3500, 1557-7295},
  doi = {10.1145/992200.992206}
}

@article{fujisawa2022magnetically,
  title = {Magnetically Confined Mountains on Accreting Neutron Stars with Multipole Magnetic Fields},
  author = {Fujisawa, Kotaro and Kisaka, Shota and Kojima, Yasufumi},
  date = {2022-09-27},
  journaltitle = {Monthly Notices of the Royal Astronomical Society},
  shortjournal = {Mon. Not. R. Astron. Soc.},
  volume = {516},
  number = {4},
  pages = {5196--5208},
  issn = {0035-8711, 1365-2966},
  doi = {10.1093/mnras/stac2585}
}

@book{jardin2010computational,
  title = {Computational {{Methods}} in {{Plasma Physics}}},
  author = {Jardin, Stephen},
  date = {2010},
  series = {Chapman and {{Hall}}/{{CRC Computational Science Ser}}},
  edition = {1st ed},
  publisher = {Taylor \& Francis Group},
  location = {Baton Rouge},
  isbn = {978-1-4398-1095-8},
  pagetotal = {1}
}

@article{jose2025hydrodynamic,
  title = {Hydrodynamic Simulations of the Recurrent Nova {{T Coronae Borealis}}: {{Nucleosynthesis}} Predictions},
  shorttitle = {Hydrodynamic Simulations of the Recurrent Nova {{T Coronae Borealis}}},
  author = {José, Jordi and Hernanz, Margarita},
  date = {2025-06},
  journaltitle = {Astronomy \& Astrophysics},
  shortjournal = {Astron. Astrophys.},
  volume = {698},
  pages = {A251},
  issn = {0004-6361, 1432-0746},
  doi = {10.1051/0004-6361/202553762}
}

@article{melatos2001hydromagnetic,
  title = {Hydromagnetic {{Structure}} of a {{Neutron Star Accreting}} at {{Its Polar Caps}}},
  author = {Melatos, A. and Phinney, E. S.},
  date = {2001},
  journaltitle = {Publications of the Astronomical Society of Australia},
  shortjournal = {Publ. Astron. Soc. Aust.},
  volume = {18},
  number = {4},
  pages = {421--430},
  issn = {1323-3580, 1448-6083},
  doi = {10.1071/AS01056}
}

@article{melatos2005gravitational,
  title = {Gravitational {{Radiation}} from an {{Accreting Millisecond Pulsar}} with a {{Magnetically Confined Mountain}}},
  author = {Melatos, A. and Payne, D. J. B.},
  date = {2005-04-20},
  journaltitle = {The Astrophysical Journal},
  shortjournal = {Astrophys. J.},
  volume = {623},
  number = {2},
  pages = {1044--1050},
  issn = {0004-637X, 1538-4357},
  doi = {10.1086/428600}
}

@article{mouschovias1974static,
  title = {Static {{Equilibria}} of the {{Interstellar Gas}} in the {{Presence}} of {{Magnetic}} and {{Gravitational Fields}}: {{Large-Scale Condensations}}},
  shorttitle = {Static {{Equilibria}} of the {{Interstellar Gas}} in the {{Presence}} of {{Magnetic}} and {{Gravitational Fields}}},
  author = {Mouschovias, Telemachos Ch.},
  date = {1974-08-01},
  journaltitle = {The Astrophysical Journal},
  shortjournal = {Astrophys. J.},
  volume = {192},
  pages = {37--50},
  issn = {0004-637X},
  doi = {10.1086/153032}
}

@article{mukherjee2012phasedependent,
  title = {A Phase-Dependent View of Cyclotron Lines from Model Accretion Mounds on Neutron Stars: {{A}} Phase-Dependent View of Cyclotron Lines},
  shorttitle = {A Phase-Dependent View of Cyclotron Lines from Model Accretion Mounds on Neutron Stars},
  author = {Mukherjee, Dipanjan and Bhattacharya, Dipankar},
  date = {2012-02-11},
  journaltitle = {Monthly Notices of the Royal Astronomical Society},
  shortjournal = {Mon. Not. R. Astron. Soc.},
  volume = {420},
  number = {1},
  pages = {720--731},
  issn = {00358711},
  doi = {10.1111/j.1365-2966.2011.20085.x}
}

@article{mukherjee2017revisiting,
  title = {Revisiting {{Field Burial}} by {{Accretion}} onto {{Neutron Stars}}},
  author = {Mukherjee, Dipanjan},
  date = {2017-09},
  journaltitle = {Journal of Astrophysics and Astronomy},
  shortjournal = {J. Astrophys. Astron.},
  volume = {38},
  number = {3},
  pages = {48},
  issn = {0250-6335, 0973-7758},
  doi = {10.1007/s12036-017-9465-6}
}

@article{payne2004burial,
  title = {Burial of the Polar Magnetic Field of an Accreting Neutron Star – {{I}}. {{Self-consistent}} Analytic and Numerical Equilibria},
  author = {Payne, D. J. B. and Melatos, A.},
  date = {2004-06-21},
  journaltitle = {Monthly Notices of the Royal Astronomical Society},
  shortjournal = {Mon. Not. R. Astron. Soc.},
  volume = {351},
  number = {2},
  pages = {569--584},
  issn = {0035-8711, 1365-2966},
  doi = {10.1111/j.1365-2966.2004.07798.x}
}

@article{priymak2011quadrupole,
  title = {Quadrupole Moment of a Magnetically Confined Mountain on an Accreting Neutron Star: Effect of the Equation of State: {{Magnetic}} Mountains: Equation of State},
  shorttitle = {Quadrupole Moment of a Magnetically Confined Mountain on an Accreting Neutron Star},
  author = {Priymak, M. and Melatos, A. and Payne, D. J. B.},
  date = {2011-11-11},
  journaltitle = {Monthly Notices of the Royal Astronomical Society},
  shortjournal = {Mon. Not. R. Astron. Soc.},
  volume = {417},
  number = {4},
  pages = {2696--2713},
  issn = {00358711},
  doi = {10.1111/j.1365-2966.2011.19431.x}
}

@article{rossetto2023magnetically,
  title = {Magnetically Confined Mountains on Accreting Neutron Stars in General Relativity},
  author = {Rossetto, Pedro H. B. and Frauendiener, Jörg and Brunet, Ryan and Melatos, Andrew},
  date = {2023-09-29},
  journaltitle = {Monthly Notices of the Royal Astronomical Society},
  shortjournal = {Mon. Not. R. Astron. Soc.},
  volume = {526},
  number = {2},
  pages = {2058--2066},
  issn = {0035-8711, 1365-2966},
  doi = {10.1093/mnras/stad2850}
}

@article{rossetto2025quadrupole,
  title = {Quadrupole {{Moment}} of a {{Magnetically Confined Mountain}} on an {{Accreting Neutron Star}} in {{General Relativity}}},
  author = {Rossetto, Pedro H. B. and Frauendiener, Jörg and Melatos, Andrew},
  date = {2025-01-20},
  journaltitle = {The Astrophysical Journal},
  shortjournal = {Astrophys. J.},
  volume = {979},
  number = {1},
  pages = {10},
  issn = {0004-637X, 1538-4357},
  doi = {10.3847/1538-4357/ada276}
}

@article{scaringi2022triggering,
  title = {Triggering Micronovae through Magnetically Confined Accretion Flows in Accreting White Dwarfs},
  author = {Scaringi, S and Groot, P J and Knigge, C and Lasota, J-P and {de~Martino}, D and Cavecchi, Y and Buckley, D A H and Camisassa, M E},
  date = {2022-05-20},
  journaltitle = {Monthly Notices of the Royal Astronomical Society: Letters},
  shortjournal = {Mon. Not. R. Astron. Soc. Lett.},
  volume = {514},
  number = {1},
  pages = {L11-L15},
  issn = {1745-3925, 1745-3933},
  doi = {10.1093/mnrasl/slac042}
}

@article{sousa2024prospects,
  title = {Prospects for the Observation of Continuous Gravitational Waves from Deformed Fast-Spinning White Dwarfs},
  author = {Sousa, Manoel F and Otoniel, Edson and Coelho, Jaziel G and {de~Araujo}, José C N},
  date = {2024-05-13},
  journaltitle = {Monthly Notices of the Royal Astronomical Society},
  shortjournal = {Mon. Not. R. Astron. Soc.},
  volume = {531},
  number = {1},
  pages = {1496--1505},
  issn = {0035-8711, 1365-2966},
  doi = {10.1093/mnras/stae1232}
}

@article{suvorov2020recycled,
  title = {Recycled Pulsars with Multipolar Magnetospheres from Accretion-Induced Magnetic Burial},
  author = {Suvorov, A G and Melatos, A},
  date = {2020-10-24},
  journaltitle = {Monthly Notices of the Royal Astronomical Society},
  shortjournal = {Mon. Not. R. Astron. Soc.},
  volume = {499},
  number = {3},
  pages = {3243--3254},
  issn = {0035-8711, 1365-2966},
  doi = {10.1093/mnras/staa3132}
}

@article{virtanen2020scipy,
  title = {{{SciPy}} 1.0: Fundamental Algorithms for Scientific Computing in {{Python}}},
  shorttitle = {{{SciPy}} 1.0},
  author = {Virtanen, Pauli and Gommers, Ralf and Oliphant, Travis E. and Haberland, Matt and Reddy, Tyler and Cournapeau, David and Burovski, Evgeni and Peterson, Pearu and Weckesser, Warren and Bright, Jonathan and Van Der Walt, Stéfan J. and Brett, Matthew and Wilson, Joshua and Millman, K. Jarrod and Mayorov, Nikolay and Nelson, Andrew R. J. and Jones, Eric and Kern, Robert and Larson, Eric and Carey, C J and Polat, İlhan and Feng, Yu and Moore, Eric W. and VanderPlas, Jake and Laxalde, Denis and Perktold, Josef and Cimrman, Robert and Henriksen, Ian and Quintero, E. A. and Harris, Charles R. and Archibald, Anne M. and Ribeiro, Antônio H. and Pedregosa, Fabian and Van Mulbregt, Paul and {SciPy 1.0 Contributors} and Vijaykumar, Aditya and Bardelli, Alessandro Pietro and Rothberg, Alex and Hilboll, Andreas and Kloeckner, Andreas and Scopatz, Anthony and Lee, Antony and Rokem, Ariel and Woods, C. Nathan and Fulton, Chad and Masson, Charles and Häggström, Christian and Fitzgerald, Clark and Nicholson, David A. and Hagen, David R. and Pasechnik, Dmitrii V. and Olivetti, Emanuele and Martin, Eric and Wieser, Eric and Silva, Fabrice and Lenders, Felix and Wilhelm, Florian and Young, G. and Price, Gavin A. and Ingold, Gert-Ludwig and Allen, Gregory E. and Lee, Gregory R. and Audren, Hervé and Probst, Irvin and Dietrich, Jörg P. and Silterra, Jacob and Webber, James T and Slavič, Janko and Nothman, Joel and Buchner, Johannes and Kulick, Johannes and Schönberger, Johannes L. and De Miranda Cardoso, José Vinícius and Reimer, Joscha and Harrington, Joseph and Rodríguez, Juan Luis Cano and Nunez-Iglesias, Juan and Kuczynski, Justin and Tritz, Kevin and Thoma, Martin and Newville, Matthew and Kümmerer, Matthias and Bolingbroke, Maximilian and Tartre, Michael and Pak, Mikhail and Smith, Nathaniel J. and Nowaczyk, Nikolai and Shebanov, Nikolay and Pavlyk, Oleksandr and Brodtkorb, Per A. and Lee, Perry and McGibbon, Robert T. and Feldbauer, Roman and Lewis, Sam and Tygier, Sam and Sievert, Scott and Vigna, Sebastiano and Peterson, Stefan and More, Surhud and Pudlik, Tadeusz and Oshima, Takuya and Pingel, Thomas J. and Robitaille, Thomas P. and Spura, Thomas and Jones, Thouis R. and Cera, Tim and Leslie, Tim and Zito, Tiziano and Krauss, Tom and Upadhyay, Utkarsh and Halchenko, Yaroslav O. and Vázquez-Baeza, Yoshiki},
  date = {2020-03-02},
  journaltitle = {Nature Methods},
  shortjournal = {Nat. Methods},
  volume = {17},
  number = {3},
  pages = {261--272},
  issn = {1548-7091, 1548-7105},
  doi = {10.1038/s41592-019-0686-2}
}

@article{wette2010sinking,
  title = {Sinking of a Magnetically Confined Mountain on an Accreting Neutron Star},
  author = {Wette, K. and Vigelius, M. and Melatos, A.},
  date = {2010-02-21},
  journaltitle = {Monthly Notices of the Royal Astronomical Society},
  shortjournal = {Mon. Not. R. Astron. Soc.},
  volume = {402},
  number = {2},
  pages = {1099--1110},
  issn = {00358711, 13652966},
  doi = {10.1111/j.1365-2966.2009.15937.x}
}

@article{yeole2025investigating,
  title = {Investigating Field Burial by Magnetically Confined Accretion Mounds on Neutron Stars},
  author = {Yeole, Saurabh and Mukherjee, Dipanjan and Mandal, Ankush},
  date = {2025-07-26},
  journaltitle = {Monthly Notices of the Royal Astronomical Society},
  shortjournal = {Mon. Not. R. Astron. Soc.},
  volume = {541},
  number = {4},
  pages = {3280--3306},
  issn = {0035-8711, 1365-2966},
  doi = {10.1093/mnras/staf1145}
}

@article{Pringle1972,
  author  = {Pringle, J. E. and Rees, M. J.},
  title   = {Accretion Disc Models for Compact X-Ray Sources},
  journal = {A\&A},
  volume  = {21},
  pages   = {1--9},
  year    = {1972}
}

@article{Elsner1977,
  author  = {Elsner, R. F. and Lamb, F. K.},
  title   = {Accretion by Magnetic Neutron Stars},
  journal = {ApJ},
  volume  = {215},
  pages   = {897--913},
  year    = {1977},
  doi     = {10.1086/155431}
}

@article{Ghosh1978,
  author  = {Ghosh, P. and Lamb, F. K.},
  title   = {Accretion by Rotating Magnetic Neutron Stars. I. Flow of Matter outside the Magnetosphere},
  journal = {ApJ},
  volume  = {223},
  pages   = {L83--L87},
  year    = {1978},
  doi     = {10.1086/182734}
}

@article{Ghosh1979,
  author  = {Ghosh, P. and Lamb, F. K.},
  title   = {Accretion by Rotating Magnetic Neutron Stars. II. Radial and Vertical Structure of the Transition Zone},
  journal = {ApJ},
  volume  = {232},
  pages   = {259--276},
  year    = {1979},
  doi     = {10.1086/157285}
}

@book{Frank2002,
  author    = {Frank, J. and King, A. and Raine, D. J.},
  title     = {Accretion Power in Astrophysics},
  edition   = {3},
  publisher = {Cambridge University Press},
  address   = {Cambridge},
  year      = {2002}
}

@article{Cropper1990,
  author  = {Cropper, M.},
  title   = {The Polars},
  journal = {Space Science Reviews},
  volume  = {54},
  pages   = {195--295},
  year    = {1990},
  doi     = {10.1007/BF00177799}
}

@book{Warner1995,
  author    = {Warner, B.},
  title     = {Cataclysmic Variable Stars},
  publisher = {Cambridge University Press},
  address   = {Cambridge},
  year      = {1995}
}

@article{Mukai2017,
  author  = {Mukai, K.},
  title   = {X-ray Emission from Accreting White Dwarfs},
  journal = {PASP},
  volume  = {129},
  pages   = {062001},
  year    = {2017},
  doi     = {10.1088/1538-3873/aa6736}
}

@article{Wickramasinghe2000,
  author  = {Wickramasinghe, D. T. and Ferrario, L.},
  title   = {Magnetism in Isolated and Binary White Dwarfs},
  journal = {PASP},
  volume  = {112},
  pages   = {873--924},
  year    = {2000},
  doi     = {10.1086/316593}
}

@article{Ferrario2015,
  author  = {Ferrario, L. and de Martino, D. and G{\"a}nsicke, B. T.},
  title   = {Magnetic White Dwarfs},
  journal = {Space Science Reviews},
  volume  = {191},
  pages   = {111--169},
  year    = {2015},
  doi     = {10.1007/s11214-015-0152-0}
}

@article{Ferrario2020,
  author  = {Ferrario, L. and de Martino, D. and G{\"a}nsicke, B. T. and others},
  title   = {The Evolution of Magnetic White Dwarfs},
  journal = {Space Science Reviews},
  volume  = {216},
  pages   = {131},
  year    = {2020},
  doi     = {10.1007/s11214-020-00745-1}
}

@article{Scaringi2022,
  author  = {Scaringi, S. and others},
  title   = {Localized Thermonuclear Bursts from Accreting Magnetic White Dwarfs},
  journal = {Nature},
  volume  = {604},
  pages   = {447--450},
  year    = {2022},
  doi     = {10.1038/s41586-022-04495-6}
  }

@article{vigelius2009resistive,
  title = {Resistive Relaxation of a Magnetically Confined Mountain on an Accreting Neutron Star},
  author = {Vigelius, M. and Melatos, A.},
  date = {2009-06-01},
  journaltitle = {Monthly Notices of the Royal Astronomical Society},
  shortjournal = {Mon. Not. R. Astron. Soc.},
  volume = {395},
  number = {4},
  pages = {1985--1998},
  issn = {00358711, 13652966},
  doi = {10.1111/j.1365-2966.2009.14698.x}
}

@article{kulsrud2020anomalous,
  title = {Anomalous Diffusion across a Tera-{{Gauss}} Magnetic Field in Accreting Neutron Stars},
  author = {Kulsrud, Russell M. and Sunyaev, Rashid},
  date = {2020-12},
  journaltitle = {Journal of Plasma Physics},
  shortjournal = {J. Plasma Phys.},
  volume = {86},
  number = {6},
  pages = {905860602},
  issn = {0022-3778, 1469-7807},
  doi = {10.1017/S0022377820001026}
}

@article{suvorov2019relaxation,
    title = {Relaxation by thermal conduction of a magnetically confined mountain on an accreting neutron star},
    volume = {484},
    issn = {0035-8711, 1365-2966},
    url = {https://academic.oup.com/mnras/article/484/1/1079/5267149},
    doi = {10.1093/mnras/sty3518},
    language = {en},
    number = {1},
    urldate = {2022-06-29},
    journal = {Monthly Notices of the Royal Astronomical Society},
    author = {Suvorov, A G and Melatos, A},
    month = mar,
    year = {2019},
    pages = {1079--1099},
}

@online{dufour2017montreal,
  title = {The {{Montreal White Dwarf Database}}: {{A Tool}} for the {{Community}}},
  shorttitle = {The {{Montreal White Dwarf Database}}},
  author = {Dufour, P. and Blouin, S. and Coutu, S. and Fortin-Archambault, M. and Thibeault, C. and Bergeron, P. and Fontaine, G.},
  date = {2017-03-01},
  volume = {509},
  location = {eprint: arXiv:1610.00986},
  doi = {10.48550/arXiv.1610.00986},
  eventtitle = {20th {{European White Dwarf Workshop}}},
  pubstate = {prepublished}
}

@ARTICLE{2023ApJ...944...56A,
       author = {{Amorim}, L.~L. and {Kepler}, S.~O. and {K{\"u}lebi}, Baybars and {Jordan}, S. and {Romero}, A.~D.},
        title = "{Catalog of Magnetic White Dwarfs with Hydrogen Dominated Atmospheres}",
      journal = {The Astrophysical Journal},
         year = 2023,
        month = feb,
       volume = {944},
       number = {1},
          eid = {56},
        pages = {56},
          doi = {10.3847/1538-4357/acaf6e},
archivePrefix = {arXiv},
       eprint = {2301.08862},
 primaryClass = {astro-ph.SR},
       adsurl = {https://ui.adsabs.harvard.edu/abs/2023ApJ...944...56A}
}

@ARTICLE{2025A&A...698A.106S,
       author = {{Schwope}, Axel D.},
        title = "{PolarCat: Catalog of polars, low-accretion rate polars, and candidate objects}",
      journal = {Astronomy and Astrophysics},
         year = 2025,
        month = jun,
       volume = {698},
          eid = {A106},
        pages = {A106},
          doi = {10.1051/0004-6361/202554519},
archivePrefix = {arXiv},
       eprint = {2505.10337},
 primaryClass = {astro-ph.SR},
       adsurl = {https://ui.adsabs.harvard.edu/abs/2025A&A...698A.106S}
}

@ARTICLE{2017MNRAS.466.2855P,
       author = {{Pala}, A.~F. and {G{\"a}nsicke}, B.~T. and {Townsley}, D. and {Boyd}, D. and {Cook}, M.~J. and {De Martino}, D. and {Godon}, P. and {Haislip}, J.~B. and {Henden}, A.~A. and {Hubeny}, I. and {Ivarsen}, K.~M. and {Kafka}, S. and {Knigge}, C. and {LaCluyze}, A.~P. and {Long}, K.~S. and {Marsh}, T.~R. and {Monard}, B. and {Moore}, J.~P. and {Myers}, G. and {Nelson}, P. and {Nogami}, D. and {Oksanen}, A. and {Pickard}, R. and {Poyner}, G. and {Reichart}, D.~E. and {Rodriguez Perez}, D. and {Schreiber}, M.~R. and {Shears}, J. and {Sion}, E.~M. and {Stubbings}, R. and {Szkody}, P. and {Zorotovic}, M.},
        title = "{Effective temperatures of cataclysmic-variable white dwarfs as a probe of their evolution}",
      journal = {Monthly Notices of the Royal Astronomical Society},
         year = 2017,
        month = apr,
       volume = {466},
       number = {3},
        pages = {2855-2878},
          doi = {10.1093/mnras/stw3293},
archivePrefix = {arXiv},
       eprint = {1701.02738},
 primaryClass = {astro-ph.SR},
       adsurl = {https://ui.adsabs.harvard.edu/abs/2017MNRAS.466.2855P}
}

@ARTICLE{2005MNRAS.360.1091S,
       author = {{Saxton}, Curtis J. and {Wu}, Kinwah and {Cropper}, Mark and {Ramsay}, Gavin},
        title = "{Two-temperature accretion flows in magnetic cataclysmic variables: structures of post-shock emission regions and X-ray spectroscopy}",
      journal = {Monthly Notices of the Royal Astronomical Society},
         year = 2005,
        month = jul,
       volume = {360},
       number = {3},
        pages = {1091-1104},
          doi = {10.1111/j.1365-2966.2005.09103.x},
archivePrefix = {arXiv},
       eprint = {astro-ph/0504267},
 primaryClass = {astro-ph},
       adsurl = {https://ui.adsabs.harvard.edu/abs/2005MNRAS.360.1091S}
}

@ARTICLE{2010A&A...520A..25Y,
       author = {{Yuasa}, T. and {Nakazawa}, K. and {Makishima}, K. and {Saitou}, K. and {Ishida}, M. and {Ebisawa}, K. and {Mori}, H. and {Yamada}, S.},
        title = "{White dwarf masses in intermediate polars observed with the Suzaku satellite}",
      journal = {Astronomy and Astrophysics},
         year = 2010,
        month = sep,
       volume = {520},
          eid = {A25},
        pages = {A25},
          doi = {10.1051/0004-6361/201014542},
archivePrefix = {arXiv},
       eprint = {1006.5323},
 primaryClass = {astro-ph.HE},
       adsurl = {https://ui.adsabs.harvard.edu/abs/2010A&A...520A..25Y}
}

@article{patterson1994,
  title={The DQ Herculis stars},
  author={Patterson, Joseph},
  journal={Publications of the Astronomical Society of the Pacific},
  volume={106},
  number={697},
  pages={209},
  year={1994}
}

@ARTICLE{1995ApJ...449L.153W,
       author = {Wang, Y. -M.},
        title = "{On the torque exerted by a magnetically threaded accretion disk}",
      journal = {The Astrophysical Journal},
         year = 1995,
        month = aug,
       volume = {449},
        pages = {L153},
          doi = {10.1086/309649},
       adsurl = {https://ui.adsabs.harvard.edu/abs/1995ApJ...449L.153W}
}

@ARTICLE{2022ApJ...941...28S,
       author = {{Sousa}, M.~F. and {Coelho}, J.~G. and {de Araujo}, J.~C.~N. and {Kepler}, S.~O. and {Rueda}, J.~A.},
        title = "{The Double White Dwarf Merger Progenitors of SDSS J2211+1136 and ZTF J1901+1458}",
      journal = {The Astrophysical Journal},
         year = 2022,
        month = dec,
       volume = {941},
       number = {1},
          eid = {28},
        pages = {28},
          doi = {10.3847/1538-4357/aca015},
archivePrefix = {arXiv},
       eprint = {2208.09506},
 primaryClass = {astro-ph.SR},
       adsurl = {https://ui.adsabs.harvard.edu/abs/2022ApJ...941...28S}
}

@article{becerra2018,
  title={The Spin Evolution of Fast-rotating, Magnetized Super-Chandrasekhar White Dwarfs in the Aftermath of White Dwarf Mergers},
  author={Becerra, L and Rueda, JA and Lor{\'e}n-Aguilar, P and Garc{\'\i}a-Berro, E},
  journal={The Astrophysical Journal},
  volume={857},
  number={2},
  pages={134},
  year={2018}
}

@ARTICLE{Pelisoli2022,
       author = {Pelisoli, I. and Marsh, T.~R. and Dhillon, V.~S. and Breedt, E. and Brown, A.~J. and Dyer, M.~J. and Green, M.~J. and Kerry, P. and Littlefair, S.~P. and Parsons, S.~G. and Sahman, D.~I. and Wild, J.~F.},
        title = "{Found: a rapidly spinning white dwarf in LAMOST J024048.51+195226.9.}",
      journal = {Monthly Notices of the Royal Astronomical Society},
         year = 2022,
        month = jan,
       volume = {509},
        pages = {L31-L36},
          doi = {10.1093/mnrasl/slab116},
       adsurl = {https://ui.adsabs.harvard.edu/abs/2022MNRAS.509L..31P}
}

@ARTICLE{2003MNRAS.338.1067P,
       author = {{Pearson}, K.~J. and {Horne}, Keith and {Skidmore}, Warren},
        title = "{Fireball models for flares in AE Aquarii}",
      journal = {Monthly Notices of the Royal Astronomical Society},
         year = 2003,
        month = feb,
       volume = {338},
       number = {4},
        pages = {1067-1083},
          doi = {10.1046/j.1365-8711.2003.06079.x},
archivePrefix = {arXiv},
       eprint = {astro-ph/0211268},
 primaryClass = {astro-ph},
       adsurl = {https://ui.adsabs.harvard.edu/abs/2003MNRAS.338.1067P}
}

@ARTICLE{1999ApJ...520..276M,
       author = {Menou, Kristen and Esin, Ann A. and Narayan, Ramesh and Garcia, Michael R. and Lasota, Jean-Pierre and McClintock, Jeffrey E.},
        title = "{Black hole and neutron star transients in quiescence}",
      journal = {The Astrophysical Journal},
         year = 1999,
        month = jul,
       volume = {520},
       number = {1},
        pages = {276-291},
          doi = {10.1086/307443},
archivePrefix = {arXiv},
       eprint = {astro-ph/9810323},
 primaryClass = {astro-ph},
       adsurl = {https://ui.adsabs.harvard.edu/abs/1999ApJ...520..276M}
}

@ARTICLE{2020MNRAS.492.5949S,
       author = {{Sousa}, Manoel F. and {Coelho}, Jaziel G. and {de Araujo}, Jos{\'e} C.~N.},
        title = "{Gravitational waves from fast-spinning white dwarfs}",
      journal = {Monthly Notices of the Royal Astronomical Society},
         year = 2020,
        month = mar,
       volume = {492},
       number = {4},
        pages = {5949-5955},
          doi = {10.1093/mnras/staa205},
archivePrefix = {arXiv},
       eprint = {2001.08534},
 primaryClass = {astro-ph.SR},
       adsurl = {https://ui.adsabs.harvard.edu/abs/2020MNRAS.492.5949S}
}

@ARTICLE{2017MNRAS.467.4484F,
       author = {{Franzon}, B. and {Schramm}, S.},
        title = "{AR Scorpii and possible gravitational wave radiation from pulsar white dwarfs}",
      journal = {Monthly Notices of the Royal Astronomical Society},
         year = 2017,
        month = jun,
       volume = {467},
       number = {4},
        pages = {4484-4490},
          doi = {10.1093/mnras/stx397},
archivePrefix = {arXiv},
       eprint = {1609.00493},
 primaryClass = {astro-ph.SR},
       adsurl = {https://ui.adsabs.harvard.edu/abs/2017MNRAS.467.4484F}
}

@ARTICLE{2019MNRAS.490.2692K,
       author = {{Kalita}, Surajit and {Mukhopadhyay}, Banibrata},
        title = "{Continuous gravitational wave from magnetized white dwarfs and neutron stars: possible missions for LISA, DECIGO, BBO, ET detectors}",
      journal = {Monthly Notices of the Royal Astronomical Society},
         year = 2019,
        month = dec,
       volume = {490},
       number = {2},
        pages = {2692-2705},
          doi = {10.1093/mnras/stz2734},
archivePrefix = {arXiv},
       eprint = {1905.02730},
 primaryClass = {astro-ph.HE},
       adsurl = {https://ui.adsabs.harvard.edu/abs/2019MNRAS.490.2692K}
}

@article{AMARO/2017,
  title={Laser Interferometer Space Antenna},
  author={Amaro-Seoane, Pau and others},
  journal = {ArXiv e-prints},
  archivePrefix = "arXiv",
  eprint = {1702.00786},
  year = 2017,
  month = feb
}

@ARTICLE{2016CQGra..33c5010L,
       author = {{Luo}, Jun and {Chen}, Li-Sheng and {Duan}, Hui-Zong and {Gong}, Yun-Gui and {Hu}, Shoucun and {Ji}, Jianghui and {Liu}, Qi and {Mei}, Jianwei and {Milyukov}, Vadim and {Sazhin}, Mikhail and {Shao}, Cheng-Gang and {Toth}, Viktor T. and {Tu}, Hai-Bo and {Wang}, Yamin and {Wang}, Yan and {Yeh}, Hsien-Chi and {Zhan}, Ming-Sheng and {Zhang}, Yonghe and {Zharov}, Vladimir and {Zhou}, Ze-Bing},
        title = "{TianQin: a space-borne gravitational wave detector}",
      journal = {Classical and Quantum Gravity},
         year = 2016,
        month = feb,
       volume = {33},
       number = {3},
          eid = {035010},
        pages = {035010},
          doi = {10.1088/0264-9381/33/3/035010},
archivePrefix = {arXiv},
       eprint = {1512.02076},
 primaryClass = {astro-ph.IM},
       adsurl = {https://ui.adsabs.harvard.edu/abs/2016CQGra..33c5010L}
}

@ARTICLE{2006CQGra..23.4887H,
       author = {{Harry}, Gregory M. and {Fritschel}, Peter and {Shaddock}, Daniel A. and {Folkner}, William and {Phinney}, E. Sterl},
        title = "{Laser interferometry for the Big Bang Observer}",
      journal = {Classical and Quantum Gravity},
         year = 2006,
        month = aug,
       volume = {23},
       number = {15},
        pages = {4887-4894},
          doi = {10.1088/0264-9381/23/15/008},
       adsurl = {https://ui.adsabs.harvard.edu/abs/2006CQGra..23.4887H}
}

@ARTICLE{2006CQGra..23S.125K,
       author = {{Kawamura}, Seiji and {Nakamura}, Takashi and {Ando}, Masaki and {Seto}, Naoki and {Tsubono}, Kimio and {Numata}, Kenji and {Takahashi}, Ryuichi and {Nagano}, Shigeo and {Ishikawa}, Takehiko and {Musha}, Mitsuru and {Ueda}, Ken-ichi and {Sato}, Takashi and {Hosokawa}, Mizuhiko and {Agatsuma}, Kazuhiro and {Akutsu}, Tomotada and {Aoyanagi}, Koh-suke and {Arai}, Koji and {Araya}, Akito and {Asada}, Hideki and {Aso}, Yoichi and {Chiba}, Takeshi and {Ebisuzaki}, Toshikazu and {Eriguchi}, Yoshiharu and {Fujimoto}, Masa-Katsu and {Fukushima}, Mitsuhiro and {Futamase}, Toshifumi and {Ganzu}, Katsuhiko and {Harada}, Tomohiro and {Hashimoto}, Tatsuaki and {Hayama}, Kazuhiro and {Hikida}, Wataru and {Himemoto}, Yoshiaki and {Hirabayashi}, Hisashi and {Hiramatsu}, Takashi and {Ichiki}, Kiyotomo and {Ikegami}, Takeshi and {Inoue}, Kaiki T. and {Ioka}, Kunihito and {Ishidoshiro}, Koji and {Itoh}, Yousuke and {Kamagasako}, Shogo and {Kanda}, Nobuyuki and {Kawashima}, Nobuki and {Kirihara}, Hiroyuki and {Kiuchi}, Kenta and {Kobayashi}, Shiho and {Kohri}, Kazunori and {Kojima}, Yasufumi and {Kokeyama}, Keiko and {Kozai}, Yoshihide and {Kudoh}, Hideaki and {Kunimori}, Hiroo and {Kuroda}, Kazuaki and {Maeda}, Kei-ichi and {Matsuhara}, Hideo and {Mino}, Yasushi and {Miyakawa}, Osamu and {Miyoki}, Shinji and {Mizusawa}, Hiromi and {Morisawa}, Toshiyuki and {Mukohyama}, Shinji and {Naito}, Isao and {Nakagawa}, Noriyasu and {Nakamura}, Kouji and {Nakano}, Hiroyuki and {Nakao}, Kenichi and {Nishizawa}, Atsushi and {Niwa}, Yoshito and {Nozawa}, Choetsu and {Ohashi}, Masatake and {Ohishi}, Naoko and {Ohkawa}, Masashi and {Okutomi}, Akira and {Oohara}, Kenichi and {Sago}, Norichika and {Saijo}, Motoyuki and {Sakagami}, Masaaki and {Sakata}, Shihori and {Sasaki}, Misao and {Sato}, Shuichi and {Shibata}, Masaru and {Shinkai}, Hisaaki and {Somiya}, Kentaro and {Sotani}, Hajime and {Sugiyama}, Naoshi and {Tagoshi}, Hideyuki and {Takahashi}, Tadayuki and {Takahashi}, Hirotaka and {Takahashi}, Ryutaro and {Takano}, Tadashi and {Tanaka}, Takahiro and {Taniguchi}, Keisuke and {Taruya}, Atsushi and {Tashiro}, Hiroyuki and {Tokunari}, Masao and {Tsujikawa}, Shinji and {Tsunesada}, Yoshiki and {Yamamoto}, Kazuhiro and {Yamazaki}, Toshitaka and {Yokoyama}, Jun'ichi and {Yoo}, Chul-Moon and {Yoshida}, Shijun and {Yoshino}, Taizoh},
        title = "{The Japanese space gravitational wave antenna{\textemdash}DECIGO}",
      journal = {Classical and Quantum Gravity},
         year = 2006,
        month = apr,
       volume = {23},
       number = {8},
        pages = {S125-S131},
          doi = {10.1088/0264-9381/23/8/S17},
       adsurl = {https://ui.adsabs.harvard.edu/abs/2006CQGra..23S.125K}
}

@ARTICLE{2017PhRvD..95j9901Y,
       author = {{Yagi}, Kent and {Seto}, Naoki},
        title = "{Erratum: Detector configuration of DECIGO/BBO and identification of cosmological neutron-star binaries [Phys. Rev. D 83, 044011 (2011)]}",
      journal = {Physical Review D},
         year = 2017,
        month = may,
       volume = {95},
       number = {10},
          eid = {109901},
        pages = {109901},
          doi = {10.1103/PhysRevD.95.109901},
       adsurl = {https://ui.adsabs.harvard.edu/abs/2017PhRvD..95j9901Y}
}

@article{suleimanov2019hard,
  title = {Hard {{X-ray}} View on Intermediate Polars in the {{{\mkbibemph{Gaia}}}} Era},
  author = {Suleimanov, Valery F and Doroshenko, Victor and Werner, Klaus},
  date = {2019-01-21},
  journaltitle = {Monthly Notices of the Royal Astronomical Society},
  shortjournal = {Mon. Not. R. Astron. Soc.},
  volume = {482},
  number = {3},
  pages = {3622--3635},
  issn = {0035-8711, 1365-2966},
  doi = {10.1093/mnras/sty2952}
}

@article{vermette2023constraining,
  title = {Constraining the {{White-dwarf Mass}} and {{Magnetic Field Strength}} of a {{New Intermediate Polar}} through {{X-Ray Observations}}},
  author = {Vermette, Benjamin and Salcedo, Ciro and Mori, Kaya and Gerber, Julian and Yoon, Kyung Duk and Bridges, Gabriel and Hailey, Charles J. and Haberl, Frank and Hong, Jaesub and Grindlay, Jonathan and Ponti, Gabriele and Ramsay, Gavin},
  date = {2023-09-01},
  journaltitle = {The Astrophysical Journal},
  shortjournal = {Astrophys. J.},
  volume = {954},
  number = {2},
  pages = {138},
  issn = {0004-637X, 1538-4357},
  doi = {10.3847/1538-4357/ace90c}
}

@article{hameury1985magnetohydrostatics,
  title = {Magnetohydrostatics in the Polar Caps of Accreting Magnetized White Dwarfs},
  author = {Hameury, J. M. and Lasota, J. P.},
  date = {1985-04-01},
  journaltitle = {Astronomy and Astrophysics},
  shortjournal = {Astron. Astrophys.},
  volume = {145},
  pages = {L10-L12},
  publisher = {EDP},
  issn = {0004-6361}
}

@article{zhang2006bottom,
  title = {The Bottom Magnetic Field and Magnetosphere Evolution of Neutron Star in Low-Mass {{X-ray}} Binary},
  author = {Zhang, C. M. and Kojima, Y.},
  date = {2006-02-11},
  journaltitle = {Monthly Notices of the Royal Astronomical Society},
  shortjournal = {Mon. Not. R. Astron. Soc.},
  volume = {366},
  number = {1},
  pages = {137--143},
  issn = {0035-8711, 1365-2966},
  doi = {10.1111/j.1365-2966.2005.09802.x}
}

@article{zhang2009there,
  title = {Is There Evidence for Field Restructuring or Decay in Accreting Magnetic White Dwarfs?},
  author = {Zhang, C. M. and Wickramasinghe, D. T. and Ferrario, Lilia},
  date = {2009-08-21},
  journaltitle = {Monthly Notices of the Royal Astronomical Society},
  shortjournal = {Mon. Not. R. Astron. Soc.},
  volume = {397},
  number = {4},
  pages = {2208--2215},
  issn = {00358711, 13652966},
  doi = {10.1111/j.1365-2966.2009.15154.x}
}
